\documentclass[reprint,amsmath,amsfonts,amssymb,aps,prl,longbibliography]{revtex4-1}
\usepackage{siunitx}
\usepackage{braket}
\usepackage{bm}
\usepackage{graphicx}

\begin{document}

\title{Fiber coupled EPR-state generation using a single temporally multiplexed squeezed light source}
\author{Mikkel V. Larsen}
\email[E-mail: ]{mivila@fysik.dtu.dk}
\author{Xueshi Guo}
\author{Casper R. Breum}
\author{Jonas S. Neergaard-Nielsen}
\email[E-mail: ]{jsne@fysik.dtu.dk}
\author{Ulrik L. Andersen}
\email[E-mail: ]{ulrik.andersen@fysik.dtu.dk}
\affiliation{Center for Macroscopic Quantum States (bigQ), Department of Physics, Technical University of Denmark, Fysikvej, 2800 Kgs. Lyngby, Denmark}
\date{\today}
\begin{abstract}
A prerequisite for universal quantum computation and other large-scale quantum information processors is the careful preparation of quantum states in massive numbers or of massive dimension. For continuous variable approaches to quantum information processing (QIP), squeezed states are the natural quantum resources, but most demonstrations have been based on a limited number of squeezed states due to the experimental complexity in up-scaling. The number of physical resources can however be significantly reduced by employing the technique of temporal multiplexing. Here, we demonstrate an application to continuous variable QIP of temporal multiplexing in fiber: Using just a single source of squeezed states in combination with active optical switching and a $\SI{200}{m}$ fiber delay line, we generate fiber-coupled Einstein-Podolsky-Rosen entangled quantum states. Our demonstration is a critical enabler for the construction of an in-fiber, all-purpose quantum information processor based on a single or few squeezed state quantum resources.
\end{abstract}
\maketitle

\noindent\textbf{INTRODUCTION}\vspace{2mm}\\
The realization of quantum computation (QC) with demonstrated quantum supremacy requires a scalable platform of quantum resources \cite{nielsen00,bennett00}: Usually hundreds of logical qubits, or thousands of physical qubits, are needed to reach this longstanding goal \cite{dalzell18}. In one-way measurement based quantum computation (MBQC) \cite{gottesman99,raussendorf01}, universal computation is performed with only single-qubit projective measurements of an entangled cluster state \cite{raussendorf03}. Thereby, scalability is relaxed to the generation of a cluster state of suitable size \cite{gu09}. Cluster states of multiple modes of light are readily accessible in continuous variable optical platforms, but most demonstrations have been limited by the amount of spatial resources \cite{glockl03,su07,dong07,yukawa08,aoki09}. However, by time and frequency multiplexing with squeezed states of light, large cluster states can be deterministically generated as demonstrated with 60 frequency modes in \cite{roslund13,chen14} and $10^6$ temporal modes in \cite{yoshikawa16,yokoyama13}. This allows for excellent scalability, thereby rendering the need for spatially distributed resources unnecessary.

MBQC based on temporally encoded cluster states \cite{menicucci11,alexander18} from a single squeezed state resource \cite{andersen16} requires optical switching and passive optical storage (such as an optical delay line) in different configurations as illustrated in Fig. 1. Multiple time-synchronized squeezed states can be generated in the network illustrated in Fig. 1a, allowing 2D cluster state generation from a single squeezing source \cite{menicucci11,alexander18}. Moreover, in MBQC, sequential measurements are performed on the cluster in which each measurement strategy is adaptively changed based on previous measurement outcomes. In some cases, switching between completely different measurement schemes, e.g. homodyne detection and a non-Gaussian measurement, is required \cite{alexander18} (Fig. 1b). As an alternative to switching between Gaussian and non-Gaussian measurement schemes, one might fix the measurement setting to Gaussian homodyne detection and switch ancillary non-Gaussian states into selected modes of the cluster state \cite{alexander17} (Fig. 1c). Finally, it is possible to realize MBQC by applying optical switching in loop-based architectures  \cite{motes14, takeda17} as illustrated in Fig. 1d. No matter which of the strategies is chosen, switching and delay lines are key functionalities in managing temporal modes in optical MBQC.

\begin{figure}
	\includegraphics[width=\linewidth]{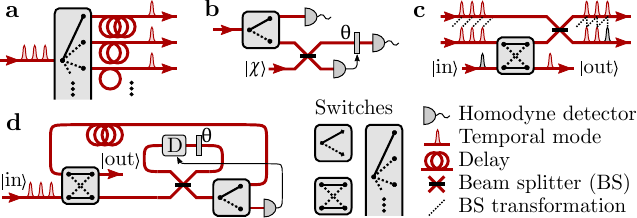}
	\caption{Quantum information processing architectures using optical switching and optical delay. (a) Switching and delay lines applied to a single squeezed state resource in order to generate multiple time-synchronized squeezed state. (b) Switching between homodyne detection and the more demanding cubic-phase gate-teleportation measurement with $\ket{\chi}$ being an ancillary cubic-phase-state \cite{alexander18}. (c) Example of switching temporal modes into or out of a cluster state \cite{alexander17}. (d) Loop-based architecture for fully temporally encoded MBQC utilizing switching and delay \cite{takeda17}.}
\end{figure}

In this article, we demonstrate optical fiber switching combined with an optical fiber delay in a continuous variable (CV) quantum setting in the telecom band. This enables us to generate an Einstein-Podolsky-Rosen (EPR) state \cite{reid09} between two fiber modes by time multiplexing of a single source of squeezed states of light. Our demonstration of optical switching and optical delay in a CV, fiber-integrated and low-loss setting is a critical step towards the realization of a scalable platform for CV quantum information processing and ultimately universal quantum computation.

\vspace{7mm}\noindent\textbf{RESULTS}\vspace{2mm}\\
The quadrature entangled EPR-state is an important resource in numerous quantum information and sensing protocols ranging from CV teleportation \cite{furusawa98} and cryptography \cite{madsen12} to CV computing \cite{menicucci06}. The most wide-spread realization of quadrature entanglement is based on cavity-enhanced spontaneous parametric down-conversion in an optical parametric oscillator (OPO). Correlations can be established between different polarization or frequency modes from a single non-degenerate OPO \cite{ou92,bardroff00,schori02,bowen02,hayasaka04,laurat05,wenger05,villar06}, or by combining the squeezed state outputs of two degenerate OPOs onto a balanced beam splitter \cite{furusawa98,silberhorn01,bowen03,takei05}. Here we use the latter approach of combining two squeezed states on a beam splitter, but instead of using two OPOs, we exploit time multiplexing of a single source.

\begin{figure}
	\includegraphics[width=\linewidth]{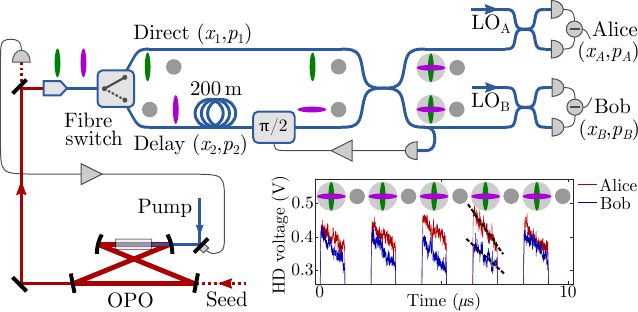}
	\caption{Schematics of the experiment. Bright amplitude squeezed states of light are generated using type-0 parametric down conversion in an optical resonator (OPO) at the wavelength of $\SI{1550}{nm}$, seeded with a coherent beam for phase locks. The squeezed states of light are coupled into a single mode fiber network (marked by blue lines) in which the generation of two-mode squeezing takes place: Using a fiber switch, two consecutive temporal modes (marked by green and purple) are guided in different directions. Subsequently, the two modes are synchronized by a fiber delay of $\SI{200}{m}$ in one of the modes. Finally, the two spatial modes interfere in a 50:50 fiber coupler, thereby forming a two-mode squeezed state in the output as the phase difference of the two input modes are locked to $\pi/2$ (using active feedback to a fiber-stretching device described in Methods). The quadratures of the two-mode squeezed state are measured with two fiber-based homodyne detection stations, Alice and Bob. A typical measurement output in time-domain is shown in the inset together with illustrations of the corresponding states in phase space, alternating between two-mode squeezed states and vacuum states.}
\end{figure}

\vspace{4mm}\noindent\textit{Experimental setup}\\
The experimental setup is sketched in Fig. 2. We inject a single ${\sim}\SI{7}{dB}$ squeezed beam into a fiber switch that alternately guides the squeezed beam into two different fibers at a frequency of $\SI{500}{kHz}$; thereby delaying one mode by $\SI{1}{\micro s}$ with respect to the other. To compensate for the delay and thus synchronize the two modes in time, the mode ahead propagates through a $\SI{200}{m}$ fiber spool. The two modes interfere with a relative phase shift of $\pi/2$ in a balanced fiber coupler, thereby forming a two-mode squeezed state.   

For state characterization, we sample on an oscilloscope the quadratures of the fiber coupler outputs measured by two homodyne detection stations, Alice and Bob. Typical time traces of such measurements are shown in the inset of Fig. 2. A single data set consists of $\SI{16000}{}$ time traces triggered by the switching signal. Each time trace is affected by a frequency dependent response of the detector giving rise to the negative slope seen in the inset of Fig. 2, and a noisy oscillatory response of the fiber switch. Besides this, there is a variation in the slope of each time trace due to spurious interferences -- both effects occur from the coherent amplitude of the initial bright squeezed state together with limited detection, switching and feedback bandwidths. However, since these effects are systematic, repeatable and synchronized with the switching process, they can be tracked and compensated in the data processing -- see Methods.

We have striven to reduce the loss of all components to maintain as much of the non-classicality as possible. We used an anti-reflection coated graded-index lens to couple the squeezed light into the fiber with an efficiency of 97\% (by matching counter-propagating light in the OPO cavity), we spliced together all fiber components to minimize fiber-to-fiber coupling losses and by using the wavelength of $\SI{1550}{nm}$, fiber propagation loss was negligible: Even through the $\SI{200}{m}$ fiber delay (standard SMF-28e+ fiber), the propagation loss is $\leq1\%$. The largest loss contribution is caused by the fiber switch (Nanona by Boston Applied Technologies Inc.), where light is coupled into a bulk electro-optic material and back into fiber leading to 17\% loss. Including OPO escape efficiency, detection efficiency and various tapping for phase locks, the total transmission from the squeezed state source to the detected signal becomes $\eta\approx68\%$ (for more details see Methods). 

\vspace{4mm}\noindent\textit{Experimental results}\\
To perform partial tomography of the generated two-mode squeezed states, we measure the quadratures $q_A(\theta)\pm q_B(-\theta)$ and $q_A(\theta)\pm q_B(\theta-\pi/2)$ as a function of the local oscillator phase $\theta$. Here $q_i(\theta)=x_i\cos\theta+p_i\sin\theta$ where $x_i$ is the amplitude and $p_i$ the phase quadrature at Alice ($i=A$) and Bob ($i=B$). The resulting noise variances at the $3$ and $\SI{10}{MHz}$ side band frequencies are shown in Fig. 3 together with theoretical predictions. We observe a maximum shot noise suppression of $\SI{3.8}{dB}$. The very small discrepancy of the measurements at $\SI{3}{MHz}$ results from technical noise of the seed beam as well as additional noise added in the delay line -- both noise effects are discussed and analyzed below and in supplementary information section 3. 

\begin{figure}
	\includegraphics[width=\linewidth]{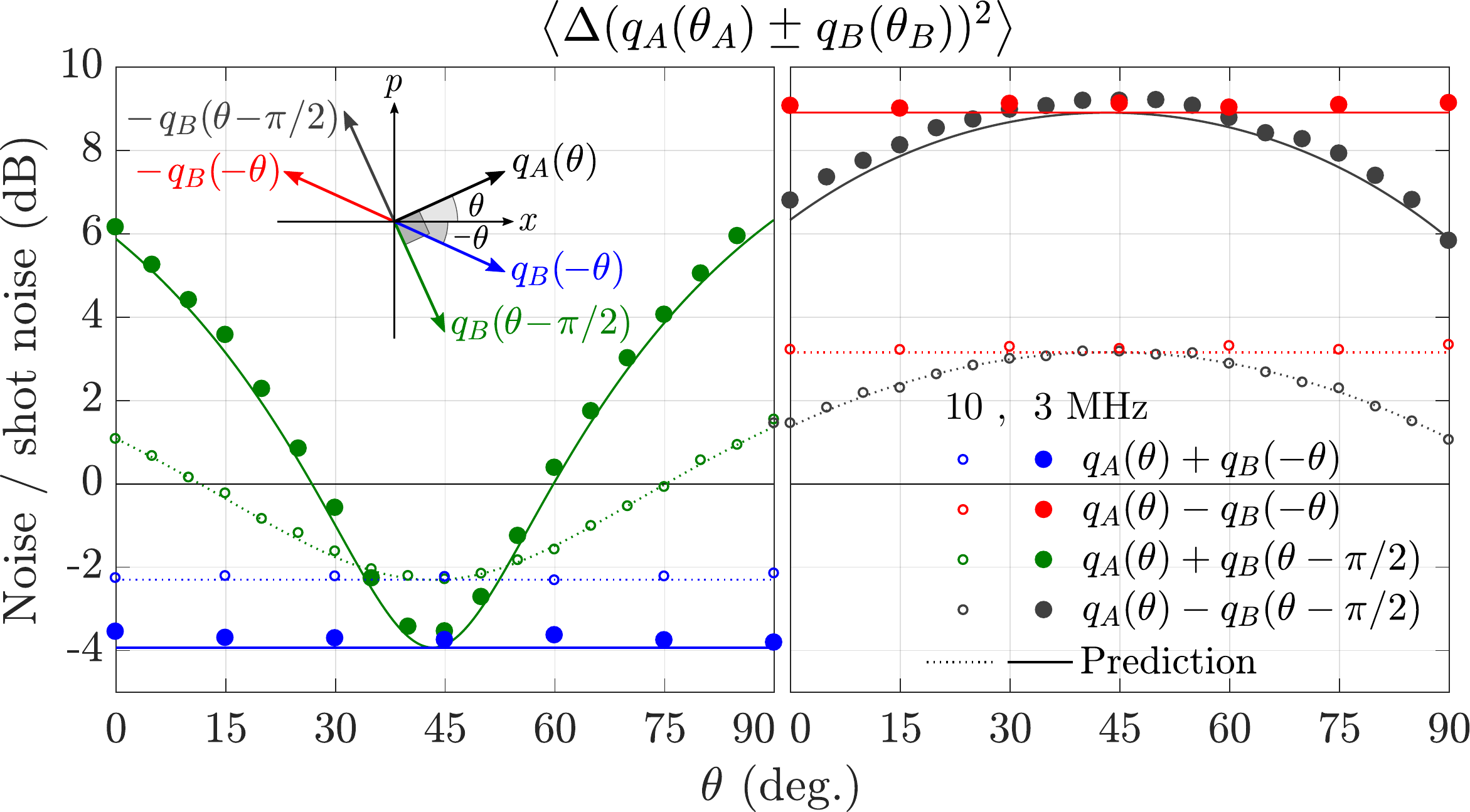}
	\caption{Partial tomography of the generated two-mode squeezed state. We plot the noise variance (normalized to the shot noise variances) of the quadratures $q_A(\theta_A)\pm q_B(\theta_B)$ with $\theta_A$ restricted to $\theta_A=\theta$ and $\theta_B$ restricted to $\theta_B=-\theta$ (blue and red), and $\theta_B=\theta-\pi/2$ (green and black). Each point corresponds to one dataset of $\SI{16000}{}$ processed time traces as in Fig. 5. To extract the 3 and $\SI{10}{MHz}$ frequency modes, each time trace is digitally mixed with a 3 or $\SI{10}{MHz}$ sine curve and integrated to one value. The noise is then the variance of these $\SI{16000}{}$ values, added/subtracted for Alice and Bob. With the time trace length of about $\SI{900}{ns}$, the frequency mode bandwidth is around $\SI{1}{MHz}$. The solid and dashed line shows theoretical noise predicted from measured efficiency, OPO bandwidth, pump power and fitted phase fluctuations in Fig. 4. The predictions include $1.7^\circ$ phase offset. The inset illustrates the measured quadratures in a phase-space diagram.}
\end{figure}

The variances of $q_A(\theta)\pm q_B(-\theta)$,
\begin{equation}\begin{split}\label{eq:qA_pm_pB}
	&\braket{\Delta(q_A\left(\theta)+q_B(-\theta)\right)^2}\\
	&\hspace{2cm}=2\left(\braket{\Delta x_1^2}\cos^2\theta+\braket{\Delta x_2^2}\sin^2\theta\right),\\
	&\braket{\Delta(q_A\left(\theta)-q_B(-\theta)\right)^2}\\
	&\hspace{2cm}=2\left(\braket{\Delta p_2^2}\cos^2\theta+\braket{\Delta p_1^2}\sin^2\theta\right),
\end{split}\end{equation}
associated with the maximally squeezed and anti-squeezed quadratures, respectively, are seen to be constant with $\theta$, indicating symmetric two-mode squeezing. This is expected as the individual single mode squeezed states in the direct ($x_1,p_1$) and delay ($x_2,p_2$) line originate from the same squeezing source, that is $\langle\Delta x_1^2\rangle=\langle\Delta x_2^2\rangle$ and $\langle\Delta p_1^2\rangle=\langle\Delta p_2^2\rangle$. From the data sets at $\theta=0^\circ$ and $90^\circ$, entanglement can be verified by the inseparability criterion \cite{duan00} which reads 
\begin{equation}
	\braket{\Delta(x_A+x_B)^2}+\braket{\Delta(p_A-p_B)^2}=1.72V_0<4V_0\;,
\end{equation}
at $\SI{3}{MHz}$, and $2.42V_0<4V_0$ at $\SI{10}{MHz}$. Here $V_0$ is the variance of the vacuum state.

When measuring the variances of
\begin{equation}\begin{split}
	&q_A(\theta)\pm q_B(\theta-\pi/2)=\,x_A\cos\theta\pm x_B\sin\theta\\
	&\hspace{4cm}+p_A\sin\theta\mp p_B\cos\theta
\end{split}\end{equation}
as a function of $\theta$, we trace out one specific projection that in particular realizes the squeezed and anti-squeezed quadratures. Maximum squeezing and anti-squeezing are measured at $\theta=45^\circ$ where correlations are strongest, corresponding to the measurements of $q_A(\theta)\pm q_B(-\theta)$. At $\theta=0^\circ$ and $90^\circ$ we expect no correlations and measure the variances $(\braket{\Delta x_1^2}+\braket{\Delta p_1^2}+\braket{\Delta x_2^2}+\braket{\Delta p_2^2})/2$ corresponding to the added noise of thermal states at Alice and Bob when tracing out one mode.

From the partial tomography, we reconstruct the covariance matrix of the two-mode squeezed state at the $\SI{3}{MHz}$ side band frequency \cite{weedbrook12}:
\begin{equation}
	\bm{\gamma}=V_0\begin{pmatrix}
		4.36 & \text{-} & -3.84 & 0.36 \\
		\text{-} & 4.43 & 0.45 & 3.92 \\
		-3.84 & 0.45 & 4.17 & \text{-} \\
		0.36 & 3.92 & \text{-} & 4.26
	\end{pmatrix}\;.
\end{equation}
Here, the entries with '-' were not measured as it would require a more elaborate measurement scheme, but they should in principle be zero due to the symmetry of the states. However, due to uncertainties in the phase control and non-perfect phase-space alignments, the values will in practice be slightly different from zero. This is also clear from the off-diagonal correlation terms $\langle x_Ap_B\rangle$ and $\langle x_Bp_A\rangle$ which in practice are non-zero as seen in the measured co-variance matrix but in theory should be zero for a perfectly aligned system (see supplementary information section 4). Finally, from the covariance matrix we determine the conditional variances between Alice and Bob's measurements from which we test the EPR-criterion \cite{reid89}:
\begin{equation}\begin{split}
	\Delta_\text{inf.}^2x_{A|B}\cdot\Delta_\text{inf.}^2p_{A|B}	&=0.69V_0^2<V_0^2\;,\\
	\Delta_\text{inf.}^2x_{B|A}\cdot\Delta_\text{inf.}^2p_{B|A}	&=0.64V_0^2<V_0^2\;,
\end{split}\end{equation}
where $\Delta_\text{inf.}^2q_{i|j} = \min_g\braket{\Delta(q_i-gq_j)^2}=\braket{\Delta q_i^2}-\braket{q_iq_j}^2/\braket{\Delta q_j^2}$ is the conditional uncertainty in predicting $q_i$ when measuring $q_j$. Since both conditional variance products are below $V_0^2$, the generated states are EPR entangled in both directions.

As seen from Eq. (\ref{eq:qA_pm_pB}) for $\theta=0^\circ$ and $90^\circ$, the measured two-mode squeezing is equivalent to the squeezing of the single mode states in the direct and delayed paths, respectively. The spectra of such measurements are shown in Fig. 4. The squeezing spectra are Lorentzian and resemble that of the OPO cavity. Furthermore, the anti-squeezing is seen to be symmetric, while the squeezing has degraded slightly in the delay line due to additional phase noise. To characterize this, we measure the seed spectrum by blocking the pump to the squeezing cavity. The low frequency noise that can be observed in the direct line results from technical noise of the seed beam. Even more low frequency noise is apparent in the squeezed state of the delay line. We believe it originates from phase noise generated by the $\SI{200}{m}$ fiber and amplitude noise from the fiber switch which is most prominent at $5\text{--}\SI{6}{MHz}$.    

\begin{figure}
	\includegraphics[width=\linewidth]{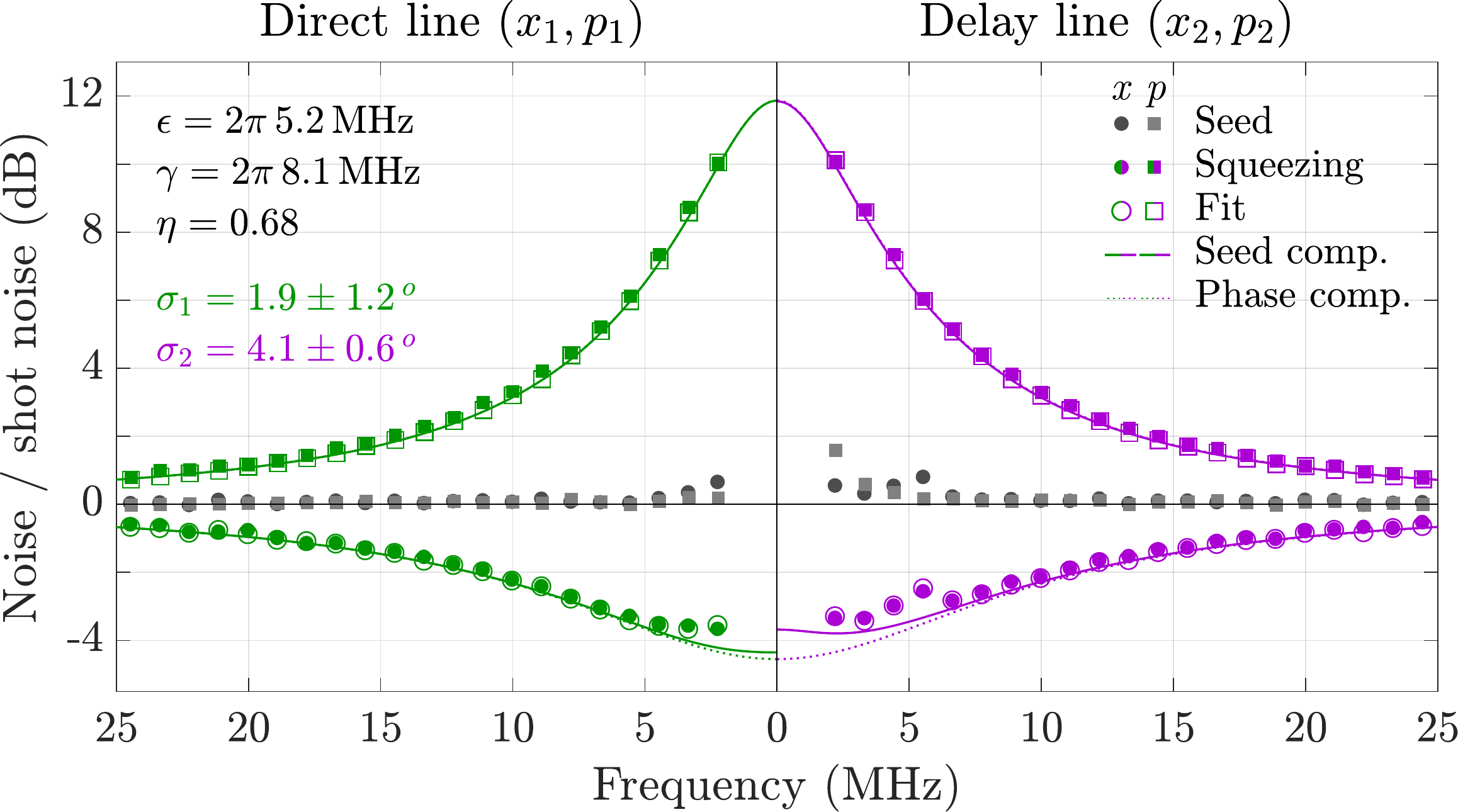}
	\caption{Spectrum of squeezing. Noise spectrum of $x_1=(x_A+x_B)/\sqrt{2}$, $p_1=(p_A+p_B)/\sqrt{2}$, $x_2=(p_A-p_B)/\sqrt{2}$ and $p_2=(x_A-x_B)/\sqrt{2}$ relative to shot noise. Solid points correspond to the average of Fourier transformed time traces in the measured datasets $q_A(\theta)\pm q_B(-\theta)$ for $\theta=0^\circ$ and $90^\circ$. Here, coloured points are with pumped OPO, while gray points are the seed noise when blocking the pump. Hollow points are the result of fitting a squeezing spectrum with phase fluctuations $\sigma_1$ and $\sigma_2$ in the direct and delay line respectively. From the fit, the solid lines indicates the expected squeezing when compensating for seed noise, while the dashed lines indicates the expected squeezing in case of no phase fluctuation, and thus the best squeezing achievable with the given efficiency.}
\end{figure}

To infer the phase fluctuations, $\sigma_i$, associated with the direct ($i=1$) and delay ($i=2$) line, the squeezing spectra including a normal distributed phase with $\sigma_i$ standard deviation, approximated to $\braket{\Delta x_i^2}\cos^2(\theta+\sigma_i)+\braket{\Delta p_i^2}\sin^2(\theta+\sigma_i)$ for $\theta=0$ and $\pi/2$ \cite{aoki06}, is fitted with $\sigma_i$ as the only fitting parameter. Here, following \cite{collett84} with additional seed noise coupled into the OPO and $V_0=1/2$, 
\begin{equation}
	\braket{\Delta q_i^2}=\frac{1}{2}\mp\frac{2\varepsilon\gamma\eta}{(\gamma\pm\varepsilon)^2+\omega^2}+\frac{K_q}{(\gamma\pm\varepsilon)^2+\omega^2},
\end{equation}
where $q=x,p$, $\varepsilon$ is the pump rate, $\gamma$ is the total OPO decay rate, $\eta$ is the overall efficiency and $\omega$ is the angular frequency, while $K_q=4\gamma\gamma_s\eta_i(\braket{\Delta q_s^2}-1/2)$ with $\gamma_s$ being the decay rate due to the seed beam coupling mirror and $\braket{\Delta q_s^2}$ is the seed beam quadrature noise before injection into the OPO (for detailed derivation see supplementary information section 3.1). We find a decay rate of $\gamma/2\pi=\SI{8.1}{MHz}$ by measuring the OPO intracavity losses (0.55\%), the cavity length ($\SI{320}{mm}$) and the transmissivity of the coupling mirror (10\%), and we estimate the pump rate to $\varepsilon/2\pi=\SI{5.2}{MHz}$ for a pump power of $\SI{350}{mW}$ and a measured OPO threshold power of $\SI{833}{mW}$. $K_q$ is estimated as $K_q=(\gamma^2+\omega^2)(\braket{\Delta q_0^2}-1/2)$ where $\braket{\Delta q_0^2}$ is the quadrature noise measured with no pump ($\varepsilon=0$, gray points in direct line of Fig. 4). Finally, to include excess noise of the delay line, the seed noise difference of the direct and delay line is added to the fit in the delay line. The fit is shown as hollow points in Fig. 4, and is seen to fit very well with the measured data. The resulting phase fluctuations obtained from the fit are $\sigma_1=1.9\pm1.2^\circ$ and $\sigma_2=4.1\pm0.6^\circ$ with uncertainties estimated as the 95\% confidence interval. These values are included in the theoretical model used for Fig. 3.

From the theoretical model with fitted phase fluctuations, the solid lines in Fig. 4 indicate the expected squeezing spectra if the seed beam were shot noise limited and no additional noise existed in the delay line. In that case, we can expect more than $\SI{4}{dB}$ two-mode squeezing. The phase fluctuation in the delay line, $\sigma_2=4.1\pm0.6^\circ$, is more than double that in the direct line, $\sigma_1=1.9\pm1.2^\circ$. This is mainly due to limited phase control bandwidth of the fiber delay and low signal-to-noise ratio of the feedback signal. Finally, the dotted line in Fig. 4 shows the squeezing spectrum we would expect if we had perfect phase control, and thus the optimum squeezing we may measure with the given efficiency.

\vspace{7mm}\noindent\textbf{DISCUSSION}\vspace{2mm}\\
The fast switching frequency of $\SI{500}{kHz}$ demonstrated here is suitable for encoding temporal modes of megahertz bandwidth and is thus applicable in the optical schemes in Fig. 1. Similarly, the low loss of the $\SI{200}{m}$ fiber allows for an efficient delay of almost $\SI{1}{\micro s}$, compatible with the temporal modes defined by the switching. However, the 17\% loss of the particular switch used here, as well as the phase fluctuations of $4^\circ$ standard deviation in the fiber delay, leads to decoherence and results in some limitations when used in quantum settings: For cluster state generation from a temporal multiplexed source, as in Fig. 1a, or when switching modes in and out of a cluster state, as in Fig. 1c, the switching loss and phase fluctuation leads to limited entanglement even when large amount of initial squeezing is available. Yet, it does not accumulate through the cluster state as the loss and phase fluctuation on each mode is local, and so it does not limit the cluster state size. It will be more detrimental in loop based architectures, as in Fig. 1d, where a temporal mode passes through the same switch and delay line multiple times, and so the switch efficiency and delay phase fluctuations limit the number of passes possible and thereby the computation depth.

High efficient fast switching is demonstrated in free-space \cite{kaneda18}, while one can imagine more compact fiber coupled switching based on Mach-Zehnder interferometry. However, in either case care must be taken not to compromise the high switching frequency, as this leads to longer delay lines necessary and thereby larger phase fluctuations. In work towards temporal encoded optical quantum information processing, faster switching is preferable as it minimizes the required delay lengths and increases the computational speed. Thus the ideal switch, besides being efficient, is as fast as the detection or squeezing source bandwidth.

In conclusion, using a single squeezing source with optical switching and delay, we have successfully generated in-fiber EPR-states with nearly $\SI{4}{dB}$ of two-mode squeezing, characterized by fiber-coupled homodyne detection. Our setup has great scalability potentials: Adding an additional delay line, it is possible to extend the setup to generate one-dimensional cluster states \cite{yoshikawa16,yokoyama13}, and by adding a multi-port switch and more delay lines,  two-dimensional cluster states \cite{menicucci11,alexander18} can be generated from a single squeezing source. Moreover, by inserting the switch inside a loop, as in Fig. 1d, combined with dynamical control, various entangled states can be generated and in principle universal quantum computation can be realized. Since all switches and delay lines are fiber components, the setup remains very small and flexible despite the increasing complexity in generating more complex states. Moreover, since fiber propagation losses are extremely low at the operating wavelength of $\SI{1550}{nm}$, decoherence is not a big issue despite the increasing number of fiber delays. The largest decoherence source in the current setup is the optical switch which introduces a loss of 17\%. However, with future developments of the optical switch, we expect that the in-fiber temporal multiplexing technique demonstrated here will play a significant role in reducing the resources in future large-scale photonic circuits for continuous variable quantum information processing, including quantum computing \cite{gu09}, quantum teleportation \cite{pirandola15}, distributed sensing \cite{guo19} and multi-partite quantum key distribution.

\vspace{7mm}\noindent\textbf{METHODS}\vspace{2mm}\\
\noindent\textit{Squeezing source}\\
The experimental setup is outlined in Fig. 2. As squeezing source, we use an optical parametric oscillator (OPO) based on a periodically poled potassium titanyl phosphate (PPKTP) crystal in a bowtie shaped cavity, locked by a counter propagating coherent beam. A pump beam at a wavelength of $\SI{775}{nm}$ is used to drive the parametric process and thus produce squeezed light at $\SI{1550}{nm}$ via type-0 phase matching. The OPO has a bandwidth of $\gamma/\pi = \SI{16}{MHz}$. Stable phase locking at different stages of the experiment is facilitated by an excitation of the squeezed state, realized by injecting a bright seed beam into the OPO. To lock the phase of the input pump beam to the deamplification point of the parametric process, thereby producing amplitude squeezed states, we tap off and detect 1\% of the excited squeezed beam for feedback to a piezo-mounted mirror in the pump beam. This, as well as all other feedback controls in the experiment, is realized by the open-source software package PyRPL \cite{neuhaus17} running on Red Pitaya boards that integrate an FPGA system-on-chip with fast ADCs and DACs.

\vspace{4mm}\noindent\textit{In-fiber phase control}\\
For locking the $\pi/2$ relative phase difference when interfering the two beams of bright squeezed states in a balanced fiber coupler for EPR-state generation, 1\% is tapped off one of the fiber coupler output arms, and fed back to a homemade fiber stretcher in the delay line based on \cite{mei07}. Here, using a piezoelectric actuator, a phase shift is induced by stretching the fiber. For more details, see the supplementary information section 2. The optical transmission efficiency is near unity, as it simply depends on the fiber which has negligible loss at $\SI{1550}{nm}$ wavelength. This allows high-efficient in-fiber phase control, and the same design is used for phase control of the local oscillators in the homodyne detection.

\vspace{4mm}\noindent\textit{Fiber-coupled homodyne detection}\\
To detect quadratures of the in-fiber generated EPR-state, we developed a fiber-coupled homodyne detector (HD) where signal and local oscillator (LO) is interfered in a balanced fiber coupler before detection. For schematics and details, see supplementary information section 2. This has the benefit of being mobile, and the visibility between signal and LO is easily optimized to near unity due to the single mode nature of the fiber used.

The fiber coupler is not exactly symmetric, but has a coupling ratio of approximately 48:52. To compensate for this, the HD is balanced by attenuation in the fiber coupler output arm of stronger LO by inducing bending losses. With an asymmetry of $4\%$ in the fiber coupler, after balancing this leads to $4\%$ loss.

Finally, to couple and focus light from the fiber onto the HD photo diodes of $\SI{100}{\micro m}$ diameter (Laser Components Nordic AB), anti-reflective coated graded-index (GRIN) lens are used in front of the diode, leading to a free-space waist diameter of $\SI{13}{\micro m}$ at $\SI{5}{mm}$ from the GRIN lens facet. The quantum efficiency is measured to be $97\%$, and so together with $4\%$ loss from balancing and 99\% visibility, the total HD efficiency achieved is 91\%.

\vspace{4mm}\noindent\textit{Overall efficiency}\\
With the OPO escape efficiency of 95\%, and 1\% tapping for gain lock, the efficiency in free-space before fiber coupling is 94\%. In fiber, including 97\% fiber coupling efficiency, 17\% loss in the fiber switch and 1\% tapping for phase control, the efficiency is 80\%. Finally, with 91\% detection efficiency, the overall efficiency becomes
\begin{equation}
    \eta=0.94\cdot0.80\cdot0.91=68\%\;.
\end{equation}

\vspace{4mm}\noindent\textit{Temporal data processing}\\
To recover two-mode squeezing from the acquired time traces affected by a frequency dependent detector response (leading to a negative slope), spurious interference (leading to slope variations) and an oscillating response from the switch, we use the statistic of $\SI{16000}{}$ time traces in a dataset synchronized with the switching process. To compensate for the negative and varying slope of each time trace, linear regression lines (as the dashed lines in the inset of Fig. 2) are subtracted from each individual trace of the dataset. The result is shown in Fig. 5(left). Here, the repeatable oscillating noise is visible, and compensated for by subtracting the average time trace of the dataset from every single time trace. The final processed dataset is seen in Fig. 5(right) with a constant temporal histogram and a single time trace at Alice and Bob showing anti-correlations as in \cite{takei06}. For detailed discussion on the data processing, see supplementary information section 2.1.

\begin{figure}
	\includegraphics[width=\linewidth]{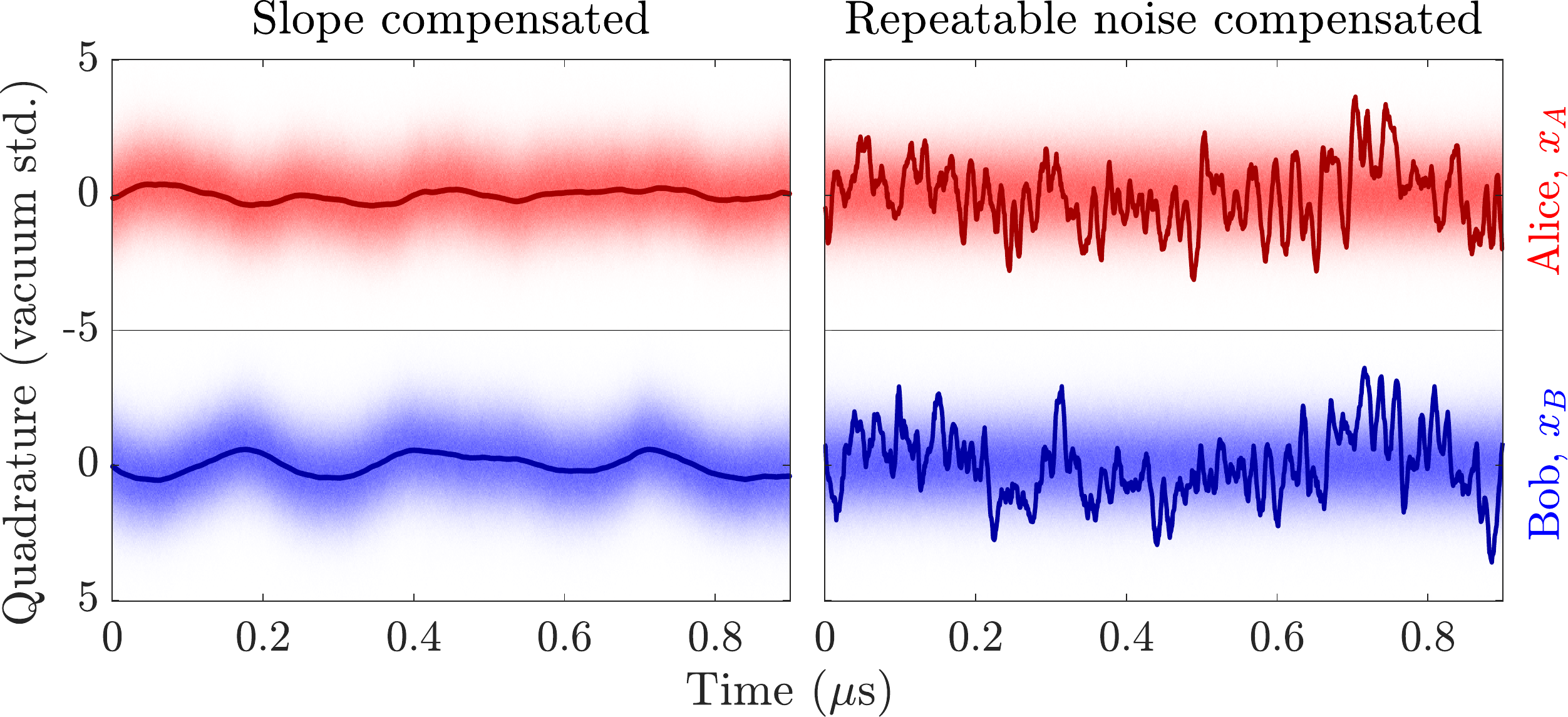}
	\caption{Temporal filtering by data processing. (left) Temporal histogram of a data set with $\SI{16000}{}$ time traces associated with amplitude quadratures of a two-mode squeezed state at Alice (red) and Bob (blue) compensated for slope variations and decaying detector response. The solid lines show the dataset average time trace indicating the remaining repeating oscillations from the switching process. (right) Temporal histogram of the dataset in (left) compensated for any systematic and repeatable noise responses from the switching process. Here, the solid lines indicate a single pair of synchronized time traces (at Alice and Bob) in which quadrature anti-correlations are visible (note the inverted axis on Bob).}
\end{figure}

\vspace{7mm}\noindent\textbf{DATA AVAILABILITY}\vspace{2mm}\\
Experimental data and analysis code is available on request.

\vspace{7mm}\noindent\textbf{ACKNOWLEDGEMENTS}\vspace{2mm}\\
The work was supported by the Danish National Research Foundation through the Center for Macroscopic Quantum States (bigQ, DNRF142), and the VILLUM FOUNDATION Young Investigator Programme.

\vspace{7mm}\noindent\textbf{AUTHOR CONTRIBUTIONS}\vspace{2mm}\\
U.L.A., J.S.N. and M.V.L. conceived the experiment. J.S.N., X.G. and C.R.B. designed and built the squeezing source. M.V.L. performed the experiment and analyzed the data. All authors contributed to the manuscript.\\


\begin{thebibliography}{49}
\makeatletter
\providecommand \@ifxundefined [1]{
 \@ifx{#1\undefined}
}
\providecommand \@ifnum [1]{
 \ifnum #1\expandafter \@firstoftwo
 \else \expandafter \@secondoftwo
 \fi
}
\providecommand \@ifx [1]{
 \ifx #1\expandafter \@firstoftwo
 \else \expandafter \@secondoftwo
 \fi
}
\providecommand \natexlab [1]{#1}%
\providecommand \enquote  [1]{``#1''}%
\providecommand \bibnamefont  [1]{#1}%
\providecommand \bibfnamefont [1]{#1}%
\providecommand \citenamefont [1]{#1}%
\providecommand \href@noop [0]{\@secondoftwo}%
\providecommand \href [0]{\begingroup \@sanitize@url \@href}%
\providecommand \@href[1]{\@@startlink{#1}\@@href}%
\providecommand \@@href[1]{\endgroup#1\@@endlink}%
\providecommand \@sanitize@url [0]{\catcode `\\12\catcode `\$12\catcode
  `\&12\catcode `\#12\catcode `\^12\catcode `\_12\catcode `\%12\relax}%
\providecommand \@@startlink[1]{}%
\providecommand \@@endlink[0]{}%
\providecommand \url  [0]{\begingroup\@sanitize@url \@url }%
\providecommand \@url [1]{\endgroup\@href {#1}{\urlprefix }}%
\providecommand \urlprefix  [0]{URL }%
\providecommand \Eprint [0]{\href }%
\providecommand \doibase [0]{http://dx.doi.org/}%
\providecommand \selectlanguage [0]{\@gobble}%
\providecommand \bibinfo  [0]{\@secondoftwo}%
\providecommand \bibfield  [0]{\@secondoftwo}%
\providecommand \translation [1]{[#1]}%
\providecommand \BibitemOpen [0]{}%
\providecommand \bibitemStop [0]{}%
\providecommand \bibitemNoStop [0]{.\EOS\space}%
\providecommand \EOS [0]{\spacefactor3000\relax}%
\providecommand \BibitemShut  [1]{\csname bibitem#1\endcsname}%
\let\auto@bib@innerbib\@empty

\bibitem [{\citenamefont {Nielsen}\ and\ \citenamefont
  {Chuang}(2000)}]{nielsen00}%
  \BibitemOpen
  \bibfield  {author} {\bibinfo {author} {\bibfnamefont {M.~A.}\ \bibnamefont
  {Nielsen}}\ and\ \bibinfo {author} {\bibfnamefont {I.~L.}\ \bibnamefont
  {Chuang}},\ }\href@noop {} {\emph {\bibinfo {title} {Quantum Computation and
  Quantum Information}}}\ (\bibinfo  {publisher} {Cambridge University Press},\
  \bibinfo {address} {England},\ \bibinfo {year} {2000})\BibitemShut {NoStop}%
\bibitem [{\citenamefont {Bennett}\ and\ \citenamefont
  {DiVincenzo}(2000)}]{bennett00}%
  \BibitemOpen
  \bibfield  {author} {\bibinfo {author} {\bibfnamefont {C.~H.}\ \bibnamefont
  {Bennett}}\ and\ \bibinfo {author} {\bibfnamefont {D.~P.}\ \bibnamefont
  {DiVincenzo}},\ }\bibfield  {title} {\enquote {\bibinfo {title} {Quantum
  information and computation},}\ }\href@noop {} {\bibfield  {journal}
  {\bibinfo  {journal} {Nature}\ }\textbf {\bibinfo {volume} {404}},\ \bibinfo
  {pages} {247} (\bibinfo {year} {2000})}\BibitemShut {NoStop}%
\bibitem [{\citenamefont {Dalzell}\ \emph {et~al.}(2018)\citenamefont
  {Dalzell}, \citenamefont {Harrow}, \citenamefont {Koh},\ and\ \citenamefont
  {Placa}}]{dalzell18}%
  \BibitemOpen
  \bibfield  {author} {\bibinfo {author} {\bibfnamefont {A.~M.}\ \bibnamefont
  {Dalzell}}, \bibinfo {author} {\bibfnamefont {A.~W.}\ \bibnamefont {Harrow}},
  \bibinfo {author} {\bibfnamefont {D.~E.}\ \bibnamefont {Koh}}, \ and\
  \bibinfo {author} {\bibfnamefont {R.~L.~La}\ \bibnamefont {Placa}},\
  }\bibfield  {title} {\enquote {\bibinfo {title} {How many qubits are needed
  for quantum computational supremacy?}}\ }\href@noop {} {\bibfield  {journal}
  {\bibinfo  {journal} {arXiv:quant-ph/1805.05224v2}\ } (\bibinfo {year}
  {2018})}\BibitemShut {NoStop}%
\bibitem [{\citenamefont {Gottesman}\ and\ \citenamefont
  {Chuang}(1999)}]{gottesman99}%
  \BibitemOpen
  \bibfield  {author} {\bibinfo {author} {\bibfnamefont {D.}~\bibnamefont
  {Gottesman}}\ and\ \bibinfo {author} {\bibfnamefont {I.~L.}\ \bibnamefont
  {Chuang}},\ }\bibfield  {title} {\enquote {\bibinfo {title} {Quantum
  teleportation as a universal computational primitive},}\ }\href@noop {}
  {\bibfield  {journal} {\bibinfo  {journal} {Nature}\ }\textbf {\bibinfo
  {volume} {402}},\ \bibinfo {pages} {390} (\bibinfo {year}
  {1999})}\BibitemShut {NoStop}%
\bibitem [{\citenamefont {Raussendorf}\ and\ \citenamefont
  {Briegel}(2001)}]{raussendorf01}%
  \BibitemOpen
  \bibfield  {author} {\bibinfo {author} {\bibfnamefont {R.}~\bibnamefont
  {Raussendorf}}\ and\ \bibinfo {author} {\bibfnamefont {H.~J.}\ \bibnamefont
  {Briegel}},\ }\bibfield  {title} {\enquote {\bibinfo {title} {A one-way
  quantum computer},}\ }\href@noop {} {\bibfield  {journal} {\bibinfo
  {journal} {Phys. Rev. Lett.}\ }\textbf {\bibinfo {volume} {86}},\ \bibinfo
  {pages} {5188} (\bibinfo {year} {2001})}\BibitemShut {NoStop}%
\bibitem [{\citenamefont {Raussendorf}\ \emph {et~al.}(2003)\citenamefont
  {Raussendorf}, \citenamefont {Browne},\ and\ \citenamefont
  {Briegel}}]{raussendorf03}%
  \BibitemOpen
  \bibfield  {author} {\bibinfo {author} {\bibfnamefont {R.}~\bibnamefont
  {Raussendorf}}, \bibinfo {author} {\bibfnamefont {D.~E.}\ \bibnamefont
  {Browne}}, \ and\ \bibinfo {author} {\bibfnamefont {H.~J.}\ \bibnamefont
  {Briegel}},\ }\bibfield  {title} {\enquote {\bibinfo {title}
  {Measurement-based quantum computation with cluster states},}\ }\href@noop {}
  {\bibfield  {journal} {\bibinfo  {journal} {Phys. Rev. A}\ }\textbf {\bibinfo
  {volume} {68}},\ \bibinfo {pages} {022312} (\bibinfo {year}
  {2003})}\BibitemShut {NoStop}%
\bibitem [{\citenamefont {Gu}\ \emph {et~al.}(2009)\citenamefont {Gu},
  \citenamefont {Weedbrook}, \citenamefont {Menicucci}, \citenamefont {Ralph},\
  and\ \citenamefont {van Loock}}]{gu09}%
  \BibitemOpen
  \bibfield  {author} {\bibinfo {author} {\bibfnamefont {M.}~\bibnamefont
  {Gu}}, \bibinfo {author} {\bibfnamefont {C.}~\bibnamefont {Weedbrook}},
  \bibinfo {author} {\bibfnamefont {N.~C.}\ \bibnamefont {Menicucci}}, \bibinfo
  {author} {\bibfnamefont {T.~C.}\ \bibnamefont {Ralph}}, \ and\ \bibinfo
  {author} {\bibfnamefont {P.}~\bibnamefont {van Loock}},\ }\bibfield  {title}
  {\enquote {\bibinfo {title} {Quantum computing with continuous-variable
  clusters},}\ }\href@noop {} {\bibfield  {journal} {\bibinfo  {journal} {Phys.
  Rev. A}\ }\textbf {\bibinfo {volume} {79}},\ \bibinfo {pages} {062318}
  (\bibinfo {year} {2009})}\BibitemShut {NoStop}%
\bibitem [{\citenamefont {et~al.}(2003)}]{glockl03}%
  \BibitemOpen
  \bibfield  {author} {\bibinfo {author} {\bibfnamefont {O.~Gl\"ockl}\
  \bibnamefont {et~al.}},\ }\bibfield  {title} {\enquote {\bibinfo {title}
  {Experiment towards continuous-variable entanglement swapping: Highly
  correlated four-partite quantum state},}\ }\href@noop {} {\bibfield
  {journal} {\bibinfo  {journal} {Phys. Rev. A}\ }\textbf {\bibinfo {volume}
  {68}},\ \bibinfo {pages} {012319} (\bibinfo {year} {2003})}\BibitemShut
  {NoStop}%
\bibitem [{\citenamefont {et~al.}(2007{\natexlab{a}})}]{su07}%
  \BibitemOpen
  \bibfield  {author} {\bibinfo {author} {\bibfnamefont {X.~Su}\ \bibnamefont
  {et~al.}},\ }\bibfield  {title} {\enquote {\bibinfo {title} {Experimental
  preparation of quadripartite cluster and {Greenberger-Horne-Zeilinger}
  entangled states for continuous variables},}\ }\href@noop {} {\bibfield
  {journal} {\bibinfo  {journal} {Phys. Rev. Lett.}\ }\textbf {\bibinfo
  {volume} {98}},\ \bibinfo {pages} {070502} (\bibinfo {year}
  {2007}{\natexlab{a}})}\BibitemShut {NoStop}%
\bibitem [{\citenamefont {et~al.}(2007{\natexlab{b}})}]{dong07}%
  \BibitemOpen
  \bibfield  {author} {\bibinfo {author} {\bibfnamefont {R.~Dong}\ \bibnamefont
  {et~al.}},\ }\bibfield  {title} {\enquote {\bibinfo {title} {An efficient
  source of continuous variable polarization entanglement},}\ }\href@noop {}
  {\bibfield  {journal} {\bibinfo  {journal} {New J. Phys.}\ }\textbf {\bibinfo
  {volume} {9}},\ \bibinfo {pages} {410} (\bibinfo {year}
  {2007}{\natexlab{b}})}\BibitemShut {NoStop}%
\bibitem [{\citenamefont {Yukawa}\ \emph {et~al.}(2008)\citenamefont {Yukawa},
  \citenamefont {Ukai}, \citenamefont {van Loock},\ and\ \citenamefont
  {Furusawa}}]{yukawa08}%
  \BibitemOpen
  \bibfield  {author} {\bibinfo {author} {\bibfnamefont {M.}~\bibnamefont
  {Yukawa}}, \bibinfo {author} {\bibfnamefont {R.}~\bibnamefont {Ukai}},
  \bibinfo {author} {\bibfnamefont {P.}~\bibnamefont {van Loock}}, \ and\
  \bibinfo {author} {\bibfnamefont {A.}~\bibnamefont {Furusawa}},\ }\bibfield
  {title} {\enquote {\bibinfo {title} {Experimental generation of four-mode
  continuous-variable cluster states},}\ }\href@noop {} {\bibfield  {journal}
  {\bibinfo  {journal} {Phys. Rev. A}\ }\textbf {\bibinfo {volume} {78}},\
  \bibinfo {pages} {012301} (\bibinfo {year} {2008})}\BibitemShut {NoStop}%
\bibitem [{\citenamefont {et~al.}(2009{\natexlab{a}})}]{aoki09}%
  \BibitemOpen
  \bibfield  {author} {\bibinfo {author} {\bibfnamefont {T.~Aoki}\ \bibnamefont
  {et~al.}},\ }\bibfield  {title} {\enquote {\bibinfo {title} {Quantum error
  correction beyond qubits},}\ }\href@noop {} {\bibfield  {journal} {\bibinfo
  {journal} {Nat. Phys.}\ }\textbf {\bibinfo {volume} {5}},\ \bibinfo {pages}
  {541} (\bibinfo {year} {2009}{\natexlab{a}})}\BibitemShut {NoStop}%
\bibitem [{\citenamefont {Roslund}\ \emph {et~al.}(2014)\citenamefont
  {Roslund}, \citenamefont {de~Ara\'ujo}, \citenamefont {Jiang}, \citenamefont
  {Fabre},\ and\ \citenamefont {Treps}}]{roslund13}%
  \BibitemOpen
  \bibfield  {author} {\bibinfo {author} {\bibfnamefont {J.}~\bibnamefont
  {Roslund}}, \bibinfo {author} {\bibfnamefont {R.~Medeiros}\ \bibnamefont
  {de~Ara\'ujo}}, \bibinfo {author} {\bibfnamefont {S.}~\bibnamefont {Jiang}},
  \bibinfo {author} {\bibfnamefont {C.}~\bibnamefont {Fabre}}, \ and\ \bibinfo
  {author} {\bibfnamefont {N.}~\bibnamefont {Treps}},\ }\bibfield  {title}
  {\enquote {\bibinfo {title} {Wavelength-multiplexed quantum networks with
  ultrafast frequency combs},}\ }\href@noop {} {\bibfield  {journal} {\bibinfo
  {journal} {Nat. Photonics}\ }\textbf {\bibinfo {volume} {8}},\ \bibinfo
  {pages} {109} (\bibinfo {year} {2014})}\BibitemShut {NoStop}%
\bibitem [{\citenamefont {Chen}\ \emph {et~al.}(2014)\citenamefont {Chen},
  \citenamefont {Menicucci},\ and\ \citenamefont {Pfister}}]{chen14}%
  \BibitemOpen
  \bibfield  {author} {\bibinfo {author} {\bibfnamefont {M.}~\bibnamefont
  {Chen}}, \bibinfo {author} {\bibfnamefont {N.~C.}\ \bibnamefont {Menicucci}},
  \ and\ \bibinfo {author} {\bibfnamefont {O.}~\bibnamefont {Pfister}},\
  }\bibfield  {title} {\enquote {\bibinfo {title} {Experimental realization of
  multipartite entanglement of 60 modes of a quantum optical frequency comb},}\
  }\href@noop {} {\bibfield  {journal} {\bibinfo  {journal} {Phys. Rev. Lett.}\
  }\textbf {\bibinfo {volume} {112}},\ \bibinfo {pages} {120505} (\bibinfo
  {year} {2014})}\BibitemShut {NoStop}%
\bibitem [{\citenamefont {et~al.}(2016)}]{yoshikawa16}%
  \BibitemOpen
  \bibfield  {author} {\bibinfo {author} {\bibfnamefont {J.~Yoshikawa}\
  \bibnamefont {et~al.}},\ }\bibfield  {title} {\enquote {\bibinfo {title}
  {Invited article: Generation of one-million-mode continuous-variable cluster
  state by unlimited time-domain multiplexing},}\ }\href@noop {} {\bibfield
  {journal} {\bibinfo  {journal} {APL Photonics}\ }\textbf {\bibinfo {volume}
  {1}},\ \bibinfo {pages} {060801} (\bibinfo {year} {2016})}\BibitemShut
  {NoStop}%
\bibitem [{\citenamefont {et~al.}(2013)}]{yokoyama13}%
  \BibitemOpen
  \bibfield  {author} {\bibinfo {author} {\bibfnamefont {S.~Yokoyama}\
  \bibnamefont {et~al.}},\ }\bibfield  {title} {\enquote {\bibinfo {title}
  {Ultra-large-scale continuous-variable cluster states multiplexed in the time
  domain},}\ }\href@noop {} {\bibfield  {journal} {\bibinfo  {journal} {Nat.
  Photonics}\ }\textbf {\bibinfo {volume} {7}},\ \bibinfo {pages} {982}
  (\bibinfo {year} {2013})}\BibitemShut {NoStop}%
\bibitem [{\citenamefont {Menicucci}(2011)}]{menicucci11}%
  \BibitemOpen
  \bibfield  {author} {\bibinfo {author} {\bibfnamefont {N.~C.}\ \bibnamefont
  {Menicucci}},\ }\bibfield  {title} {\enquote {\bibinfo {title} {Temporal-mode
  continuous-variable cluster states using linear optics},}\ }\href@noop {}
  {\bibfield  {journal} {\bibinfo  {journal} {Phys. Rev. A}\ }\textbf {\bibinfo
  {volume} {83}},\ \bibinfo {pages} {062314} (\bibinfo {year}
  {2011})}\BibitemShut {NoStop}%
\bibitem [{\citenamefont {Alexander}\ \emph {et~al.}(2018)\citenamefont
  {Alexander}, \citenamefont {Yokoyama}, \citenamefont {Furusawa},\ and\
  \citenamefont {Menicucci}}]{alexander18}%
  \BibitemOpen
  \bibfield  {author} {\bibinfo {author} {\bibfnamefont {R.~N.}\ \bibnamefont
  {Alexander}}, \bibinfo {author} {\bibfnamefont {S.}~\bibnamefont {Yokoyama}},
  \bibinfo {author} {\bibfnamefont {A.}~\bibnamefont {Furusawa}}, \ and\
  \bibinfo {author} {\bibfnamefont {N.~C.}\ \bibnamefont {Menicucci}},\
  }\bibfield  {title} {\enquote {\bibinfo {title} {Universal quantum
  computation with temporal-mode bilayer square lattices},}\ }\href@noop {}
  {\bibfield  {journal} {\bibinfo  {journal} {Phys. Rev. A}\ }\textbf {\bibinfo
  {volume} {97}},\ \bibinfo {pages} {032302} (\bibinfo {year}
  {2018})}\BibitemShut {NoStop}%
\bibitem [{\citenamefont {Andersen}\ \emph {et~al.}(2016)\citenamefont
  {Andersen}, \citenamefont {Gehring}, \citenamefont {Marquardt},\ and\
  \citenamefont {Leuchs}}]{andersen16}%
  \BibitemOpen
  \bibfield  {author} {\bibinfo {author} {\bibfnamefont {U.~L.}\ \bibnamefont
  {Andersen}}, \bibinfo {author} {\bibfnamefont {T.}~\bibnamefont {Gehring}},
  \bibinfo {author} {\bibfnamefont {C.}~\bibnamefont {Marquardt}}, \ and\
  \bibinfo {author} {\bibfnamefont {G.}~\bibnamefont {Leuchs}},\ }\bibfield
  {title} {\enquote {\bibinfo {title} {30 years of squeezed light
  generation},}\ }\href@noop {} {\bibfield  {journal} {\bibinfo  {journal}
  {Physica Scripta}\ }\textbf {\bibinfo {volume} {91}},\ \bibinfo {pages}
  {053001} (\bibinfo {year} {2016})}\BibitemShut {NoStop}%
\bibitem [{\citenamefont {Alexander}\ \emph {et~al.}(2017)\citenamefont
  {Alexander}, \citenamefont {Gabay}, \citenamefont {Rohde},\ and\
  \citenamefont {Menicucci}}]{alexander17}%
  \BibitemOpen
  \bibfield  {author} {\bibinfo {author} {\bibfnamefont {R.~N.}\ \bibnamefont
  {Alexander}}, \bibinfo {author} {\bibfnamefont {N.~C.}\ \bibnamefont
  {Gabay}}, \bibinfo {author} {\bibfnamefont {P.~P.}\ \bibnamefont {Rohde}}, \
  and\ \bibinfo {author} {\bibfnamefont {N.~C.}\ \bibnamefont {Menicucci}},\
  }\bibfield  {title} {\enquote {\bibinfo {title} {Measurement-based linear
  optics},}\ }\href@noop {} {\bibfield  {journal} {\bibinfo  {journal} {Phys.
  Rev. Lett.}\ }\textbf {\bibinfo {volume} {118}},\ \bibinfo {pages} {110503}
  (\bibinfo {year} {2017})}\BibitemShut {NoStop}%
\bibitem [{\citenamefont {Motes}\ \emph {et~al.}(2014)\citenamefont {Motes},
  \citenamefont {Gilchrist}, \citenamefont {Dowling},\ and\ \citenamefont
  {Rohde}}]{motes14}%
  \BibitemOpen
  \bibfield  {author} {\bibinfo {author} {\bibfnamefont {K.~R.}\ \bibnamefont
  {Motes}}, \bibinfo {author} {\bibfnamefont {A.}~\bibnamefont {Gilchrist}},
  \bibinfo {author} {\bibfnamefont {J.~P.}\ \bibnamefont {Dowling}}, \ and\
  \bibinfo {author} {\bibfnamefont {P.~P.}\ \bibnamefont {Rohde}},\ }\bibfield
  {title} {\enquote {\bibinfo {title} {Scalable boson sampling with time-bin
  encoding using a loop-based architecture},}\ }\href@noop {} {\bibfield
  {journal} {\bibinfo  {journal} {Phys. Rev. Lett.}\ }\textbf {\bibinfo
  {volume} {113}},\ \bibinfo {pages} {120501} (\bibinfo {year}
  {2014})}\BibitemShut {NoStop}%
\bibitem [{\citenamefont {Takeda}\ and\ \citenamefont
  {Furusawa}(2017)}]{takeda17}%
  \BibitemOpen
  \bibfield  {author} {\bibinfo {author} {\bibfnamefont {S.}~\bibnamefont
  {Takeda}}\ and\ \bibinfo {author} {\bibfnamefont {A.}~\bibnamefont
  {Furusawa}},\ }\bibfield  {title} {\enquote {\bibinfo {title} {Universal
  quantum computing with measurement-induced continuous-variable gate sequence
  in a loop-based architecture},}\ }\href@noop {} {\bibfield  {journal}
  {\bibinfo  {journal} {Phys. Rev. Lett.}\ }\textbf {\bibinfo {volume} {119}},\
  \bibinfo {pages} {120504} (\bibinfo {year} {2017})}\BibitemShut {NoStop}%
\bibitem [{\citenamefont {et~al.}(2009{\natexlab{b}})}]{reid09}%
  \BibitemOpen
  \bibfield  {author} {\bibinfo {author} {\bibfnamefont {M.~D.~Reid}\
  \bibnamefont {et~al.}},\ }\bibfield  {title} {\enquote {\bibinfo {title}
  {Colloquium: The {Einstein-Podolsky-Rosen} paradox: From concepts to
  applications},}\ }\href@noop {} {\bibfield  {journal} {\bibinfo  {journal}
  {Rev. Mod. Phys.}\ }\textbf {\bibinfo {volume} {81}},\ \bibinfo {pages}
  {1727} (\bibinfo {year} {2009}{\natexlab{b}})}\BibitemShut {NoStop}%
\bibitem [{\citenamefont {et~al.}(1998)}]{furusawa98}%
  \BibitemOpen
  \bibfield  {author} {\bibinfo {author} {\bibfnamefont {A.~Furusawa}\
  \bibnamefont {et~al.}},\ }\bibfield  {title} {\enquote {\bibinfo {title}
  {Unconditional quantum teleportation},}\ }\href@noop {} {\bibfield  {journal}
  {\bibinfo  {journal} {Science}\ }\textbf {\bibinfo {volume} {282}},\ \bibinfo
  {pages} {706} (\bibinfo {year} {1998})}\BibitemShut {NoStop}%
\bibitem [{\citenamefont {Madsen}\ \emph {et~al.}(2012)\citenamefont {Madsen},
  \citenamefont {Usenko}, \citenamefont {Lassen}, \citenamefont {Filip},\ and\
  \citenamefont {Andersen}}]{madsen12}%
  \BibitemOpen
  \bibfield  {author} {\bibinfo {author} {\bibfnamefont {L.~S.}\ \bibnamefont
  {Madsen}}, \bibinfo {author} {\bibfnamefont {V.~C.}\ \bibnamefont {Usenko}},
  \bibinfo {author} {\bibfnamefont {M.}~\bibnamefont {Lassen}}, \bibinfo
  {author} {\bibfnamefont {R.}~\bibnamefont {Filip}}, \ and\ \bibinfo {author}
  {\bibfnamefont {U.~L.}\ \bibnamefont {Andersen}},\ }\bibfield  {title}
  {\enquote {\bibinfo {title} {Continuous variable quantum key distribution
  with modulated entangled states},}\ }\href@noop {} {\bibfield  {journal}
  {\bibinfo  {journal} {Nat. Commun.}\ }\textbf {\bibinfo {volume} {3}},\
  \bibinfo {pages} {1083} (\bibinfo {year} {2012})}\BibitemShut {NoStop}%
\bibitem [{\citenamefont {et~al.}(2006)}]{menicucci06}%
  \BibitemOpen
  \bibfield  {author} {\bibinfo {author} {\bibfnamefont {N.~C.~Menicucci}\
  \bibnamefont {et~al.}},\ }\bibfield  {title} {\enquote {\bibinfo {title}
  {Universal quantum computation with continuous-variable cluster states},}\
  }\href@noop {} {\bibfield  {journal} {\bibinfo  {journal} {Phys. Rev. Lett.}\
  }\textbf {\bibinfo {volume} {97}},\ \bibinfo {pages} {110501} (\bibinfo
  {year} {2006})}\BibitemShut {NoStop}%
\bibitem [{\citenamefont {Ou}\ \emph {et~al.}(1992)\citenamefont {Ou},
  \citenamefont {Pereira}, \citenamefont {Kimble},\ and\ \citenamefont
  {Peng}}]{ou92}%
  \BibitemOpen
  \bibfield  {author} {\bibinfo {author} {\bibfnamefont {Z.~Y.}\ \bibnamefont
  {Ou}}, \bibinfo {author} {\bibfnamefont {S.~F.}\ \bibnamefont {Pereira}},
  \bibinfo {author} {\bibfnamefont {H.~J.}\ \bibnamefont {Kimble}}, \ and\
  \bibinfo {author} {\bibfnamefont {K.~C.}\ \bibnamefont {Peng}},\ }\bibfield
  {title} {\enquote {\bibinfo {title} {Realization of the
  {Einstein-Podolsky-Rosen} paradox for continuous variable},}\ }\href@noop {}
  {\bibfield  {journal} {\bibinfo  {journal} {Phys. Rev. Lett.}\ }\textbf
  {\bibinfo {volume} {68}},\ \bibinfo {pages} {3663} (\bibinfo {year}
  {1992})}\BibitemShut {NoStop}%
\bibitem [{\citenamefont {Bardroff}\ and\ \citenamefont
  {Stenholm}(2000)}]{bardroff00}%
  \BibitemOpen
  \bibfield  {author} {\bibinfo {author} {\bibfnamefont {P.~J.}\ \bibnamefont
  {Bardroff}}\ and\ \bibinfo {author} {\bibfnamefont {S.}~\bibnamefont
  {Stenholm}},\ }\bibfield  {title} {\enquote {\bibinfo {title} {Two-mode laser
  with excess noise},}\ }\href@noop {} {\bibfield  {journal} {\bibinfo
  {journal} {Phys. Rev. A}\ }\textbf {\bibinfo {volume} {62}},\ \bibinfo
  {pages} {023814} (\bibinfo {year} {2000})}\BibitemShut {NoStop}%
\bibitem [{\citenamefont {Schori}\ \emph {et~al.}(2002)\citenamefont {Schori},
  \citenamefont {S{\o}rensen},\ and\ \citenamefont {Polzik}}]{schori02}%
  \BibitemOpen
  \bibfield  {author} {\bibinfo {author} {\bibfnamefont {C.}~\bibnamefont
  {Schori}}, \bibinfo {author} {\bibfnamefont {J.~L.}\ \bibnamefont
  {S{\o}rensen}}, \ and\ \bibinfo {author} {\bibfnamefont {E.~S.}\ \bibnamefont
  {Polzik}},\ }\bibfield  {title} {\enquote {\bibinfo {title} {Narrow-band
  frequency tunable light source of continuous quadrature entanglement},}\
  }\href@noop {} {\bibfield  {journal} {\bibinfo  {journal} {Phys. Rev. A}\
  }\textbf {\bibinfo {volume} {66}},\ \bibinfo {pages} {033802} (\bibinfo
  {year} {2002})}\BibitemShut {NoStop}%
\bibitem [{\citenamefont {Bowen}\ \emph {et~al.}(2002)\citenamefont {Bowen},
  \citenamefont {Treps}, \citenamefont {Schnabel},\ and\ \citenamefont
  {Lam}}]{bowen02}%
  \BibitemOpen
  \bibfield  {author} {\bibinfo {author} {\bibfnamefont {W.~P.}\ \bibnamefont
  {Bowen}}, \bibinfo {author} {\bibfnamefont {N.}~\bibnamefont {Treps}},
  \bibinfo {author} {\bibfnamefont {R.}~\bibnamefont {Schnabel}}, \ and\
  \bibinfo {author} {\bibfnamefont {P.~K.}\ \bibnamefont {Lam}},\ }\bibfield
  {title} {\enquote {\bibinfo {title} {Experimental demonstration of continuous
  variable polarization entanglement},}\ }\href@noop {} {\bibfield  {journal}
  {\bibinfo  {journal} {Phys. Rev. Lett.}\ }\textbf {\bibinfo {volume} {89}},\
  \bibinfo {pages} {253601} (\bibinfo {year} {2002})}\BibitemShut {NoStop}%
\bibitem [{\citenamefont {Hayasaka}\ \emph {et~al.}(2004)\citenamefont
  {Hayasaka}, \citenamefont {Zhang},\ and\ \citenamefont {Kasai}}]{hayasaka04}%
  \BibitemOpen
  \bibfield  {author} {\bibinfo {author} {\bibfnamefont {K.}~\bibnamefont
  {Hayasaka}}, \bibinfo {author} {\bibfnamefont {Y.}~\bibnamefont {Zhang}}, \
  and\ \bibinfo {author} {\bibfnamefont {K.}~\bibnamefont {Kasai}},\ }\bibfield
   {title} {\enquote {\bibinfo {title} {Generation of twin beams from an
  optical parametric oscillator pumped by a frequency-doubled diode laser},}\
  }\href@noop {} {\bibfield  {journal} {\bibinfo  {journal} {Optics Lett.}\
  }\textbf {\bibinfo {volume} {29}},\ \bibinfo {pages} {1665} (\bibinfo {year}
  {2004})}\BibitemShut {NoStop}%
\bibitem [{\citenamefont {Laurat}\ \emph {et~al.}(2005)\citenamefont {Laurat},
  \citenamefont {Coudreau}, \citenamefont {Keller}, \citenamefont {Treps},\
  and\ \citenamefont {Fabre}}]{laurat05}%
  \BibitemOpen
  \bibfield  {author} {\bibinfo {author} {\bibfnamefont {J.}~\bibnamefont
  {Laurat}}, \bibinfo {author} {\bibfnamefont {T.}~\bibnamefont {Coudreau}},
  \bibinfo {author} {\bibfnamefont {G.}~\bibnamefont {Keller}}, \bibinfo
  {author} {\bibfnamefont {N.}~\bibnamefont {Treps}}, \ and\ \bibinfo {author}
  {\bibfnamefont {C.}~\bibnamefont {Fabre}},\ }\bibfield  {title} {\enquote
  {\bibinfo {title} {Effects of mode coupling on the generation of quadrature
  {Einstein-Podolsky-Rosen} entanglement in a type-{II} optical parametric
  oscillator below threshold},}\ }\href@noop {} {\bibfield  {journal} {\bibinfo
   {journal} {Phys. Rev. A}\ }\textbf {\bibinfo {volume} {71}},\ \bibinfo
  {pages} {022313} (\bibinfo {year} {2005})}\BibitemShut {NoStop}%
\bibitem [{\citenamefont {Wenger}\ \emph {et~al.}(2005)\citenamefont {Wenger},
  \citenamefont {Ourjoumtsev}, \citenamefont {Tualle-Brouri},\ and\
  \citenamefont {Grangier}}]{wenger05}%
  \BibitemOpen
  \bibfield  {author} {\bibinfo {author} {\bibfnamefont {J.}~\bibnamefont
  {Wenger}}, \bibinfo {author} {\bibfnamefont {A.}~\bibnamefont {Ourjoumtsev}},
  \bibinfo {author} {\bibfnamefont {R.}~\bibnamefont {Tualle-Brouri}}, \ and\
  \bibinfo {author} {\bibfnamefont {P.}~\bibnamefont {Grangier}},\ }\bibfield
  {title} {\enquote {\bibinfo {title} {Time-resolved homodyne characterization
  of individual quadrature-entangled pulses},}\ }\href@noop {} {\bibfield
  {journal} {\bibinfo  {journal} {Eur. Phys. J. D}\ }\textbf {\bibinfo {volume}
  {32}},\ \bibinfo {pages} {391} (\bibinfo {year} {2005})}\BibitemShut
  {NoStop}%
\bibitem [{\citenamefont {Villar}\ \emph {et~al.}(2006)\citenamefont {Villar},
  \citenamefont {Martinelli}, \citenamefont {Fabre},\ and\ \citenamefont
  {Nussenzveig}}]{villar06}%
  \BibitemOpen
  \bibfield  {author} {\bibinfo {author} {\bibfnamefont {A.~S.}\ \bibnamefont
  {Villar}}, \bibinfo {author} {\bibfnamefont {M.}~\bibnamefont {Martinelli}},
  \bibinfo {author} {\bibfnamefont {C.}~\bibnamefont {Fabre}}, \ and\ \bibinfo
  {author} {\bibfnamefont {P.}~\bibnamefont {Nussenzveig}},\ }\bibfield
  {title} {\enquote {\bibinfo {title} {Direct production of tripartite
  pump-signal-idler entanglement in the above-threshold optical parametric
  oscillator},}\ }\href@noop {} {\bibfield  {journal} {\bibinfo  {journal}
  {Phys. Rev. Lett.}\ }\textbf {\bibinfo {volume} {97}},\ \bibinfo {pages}
  {140504} (\bibinfo {year} {2006})}\BibitemShut {NoStop}%
\bibitem [{\citenamefont {et~al.}(2001)}]{silberhorn01}%
  \BibitemOpen
  \bibfield  {author} {\bibinfo {author} {\bibfnamefont {C.~Silberhorn}\
  \bibnamefont {et~al.}},\ }\bibfield  {title} {\enquote {\bibinfo {title}
  {Generation of continuous variable {Einstein-Podolsky-Rosen} entanglement via
  the {Kerr} nonlinearity in an optical fiber},}\ }\href@noop {} {\bibfield
  {journal} {\bibinfo  {journal} {Phys. Rev. Lett.}\ }\textbf {\bibinfo
  {volume} {86}},\ \bibinfo {pages} {4267} (\bibinfo {year}
  {2001})}\BibitemShut {NoStop}%
\bibitem [{\citenamefont {Bowen}\ \emph {et~al.}(2003)\citenamefont {Bowen},
  \citenamefont {Schnabel}, \citenamefont {Lam},\ and\ \citenamefont
  {Ralph}}]{bowen03}%
  \BibitemOpen
  \bibfield  {author} {\bibinfo {author} {\bibfnamefont {W.~P.}\ \bibnamefont
  {Bowen}}, \bibinfo {author} {\bibfnamefont {R.}~\bibnamefont {Schnabel}},
  \bibinfo {author} {\bibfnamefont {P.~K.}\ \bibnamefont {Lam}}, \ and\
  \bibinfo {author} {\bibfnamefont {T.~C.}\ \bibnamefont {Ralph}},\ }\bibfield
  {title} {\enquote {\bibinfo {title} {Experimental investigation of criteria
  for continuous variable entanglement},}\ }\href@noop {} {\bibfield  {journal}
  {\bibinfo  {journal} {Phys. Rev. Lett.}\ }\textbf {\bibinfo {volume} {90}},\
  \bibinfo {pages} {043601} (\bibinfo {year} {2003})}\BibitemShut {NoStop}%
\bibitem [{\citenamefont {Takei}\ \emph {et~al.}(2005)\citenamefont {Takei},
  \citenamefont {Yonezawa}, \citenamefont {Aoki},\ and\ \citenamefont
  {Furusawa}}]{takei05}%
  \BibitemOpen
  \bibfield  {author} {\bibinfo {author} {\bibfnamefont {N.}~\bibnamefont
  {Takei}}, \bibinfo {author} {\bibfnamefont {H.}~\bibnamefont {Yonezawa}},
  \bibinfo {author} {\bibfnamefont {T.}~\bibnamefont {Aoki}}, \ and\ \bibinfo
  {author} {\bibfnamefont {A.}~\bibnamefont {Furusawa}},\ }\bibfield  {title}
  {\enquote {\bibinfo {title} {High-fidelity teleportation beyond the
  no-cloning limit and entanglement swapping for continuous variables},}\
  }\href@noop {} {\bibfield  {journal} {\bibinfo  {journal} {Phys. Rev. Lett.}\
  }\textbf {\bibinfo {volume} {94}},\ \bibinfo {pages} {220502} (\bibinfo
  {year} {2005})}\BibitemShut {NoStop}%
\bibitem [{\citenamefont {Duan}\ \emph {et~al.}(2000)\citenamefont {Duan},
  \citenamefont {Giedke}, \citenamefont {Cirac},\ and\ \citenamefont
  {Zoller}}]{duan00}%
  \BibitemOpen
\bibfield  {journal} {  }\bibfield  {author} {\bibinfo {author} {\bibfnamefont
  {L.}~\bibnamefont {Duan}}, \bibinfo {author} {\bibfnamefont {G.}~\bibnamefont
  {Giedke}}, \bibinfo {author} {\bibfnamefont {J.~I.}\ \bibnamefont {Cirac}}, \
  and\ \bibinfo {author} {\bibfnamefont {P.}~\bibnamefont {Zoller}},\
  }\bibfield  {title} {\enquote {\bibinfo {title} {Inseparability criterion for
  continuous variable systems},}\ }\href@noop {} {\bibfield  {journal}
  {\bibinfo  {journal} {Phys. Rev. Lett.}\ }\textbf {\bibinfo {volume} {84}},\
  \bibinfo {pages} {2722} (\bibinfo {year} {2000})}\BibitemShut {NoStop}%
\bibitem [{\citenamefont {et~al.}(2012)}]{weedbrook12}%
  \BibitemOpen
  \bibfield  {author} {\bibinfo {author} {\bibfnamefont {C.~Weedbrook}\
  \bibnamefont {et~al.}},\ }\bibfield  {title} {\enquote {\bibinfo {title}
  {Gaussian quantum information},}\ }\href@noop {} {\bibfield  {journal}
  {\bibinfo  {journal} {Rev. Mod. Phys.}\ }\textbf {\bibinfo {volume} {84}},\
  \bibinfo {pages} {621} (\bibinfo {year} {2012})}\BibitemShut {NoStop}%
\bibitem [{\citenamefont {Reid}(1989)}]{reid89}%
  \BibitemOpen
  \bibfield  {author} {\bibinfo {author} {\bibfnamefont {M.~D.}\ \bibnamefont
  {Reid}},\ }\bibfield  {title} {\enquote {\bibinfo {title} {Demonstration of
  the {Einstein-Podolsky-Rosen} paradox using nondegenerate parametric
  amplification},}\ }\href@noop {} {\bibfield  {journal} {\bibinfo  {journal}
  {Phys. Rev. A}\ }\textbf {\bibinfo {volume} {40}},\ \bibinfo {pages} {913}
  (\bibinfo {year} {1989})}\BibitemShut {NoStop}%
\bibitem [{\citenamefont {Aoki}\ \emph {et~al.}(2006)\citenamefont {Aoki},
  \citenamefont {Takahashi},\ and\ \citenamefont {Furusawa}}]{aoki06}%
  \BibitemOpen
  \bibfield  {author} {\bibinfo {author} {\bibfnamefont {T.}~\bibnamefont
  {Aoki}}, \bibinfo {author} {\bibfnamefont {G.}~\bibnamefont {Takahashi}}, \
  and\ \bibinfo {author} {\bibfnamefont {A.}~\bibnamefont {Furusawa}},\
  }\bibfield  {title} {\enquote {\bibinfo {title} {Squeezing at 946nm with
  periodically poled {KTiOPO4}},}\ }\href@noop {} {\bibfield  {journal}
  {\bibinfo  {journal} {Opt. Express}\ }\textbf {\bibinfo {volume} {14}},\
  \bibinfo {pages} {6930} (\bibinfo {year} {2006})}\BibitemShut {NoStop}%
\bibitem [{\citenamefont {Collett}\ and\ \citenamefont
  {Gardiner}(1984)}]{collett84}%
  \BibitemOpen
  \bibfield  {author} {\bibinfo {author} {\bibfnamefont {M.~J.}\ \bibnamefont
  {Collett}}\ and\ \bibinfo {author} {\bibfnamefont {C.~W.}\ \bibnamefont
  {Gardiner}},\ }\bibfield  {title} {\enquote {\bibinfo {title} {Squeezing of
  intracavity and traveling-wave light fields produced in parametric
  amplification},}\ }\href@noop {} {\bibfield  {journal} {\bibinfo  {journal}
  {Phys. Rev. A}\ }\textbf {\bibinfo {volume} {30}},\ \bibinfo {pages} {1386}
  (\bibinfo {year} {1984})}\BibitemShut {NoStop}%
\bibitem [{\citenamefont {Kaneda}\ and\ \citenamefont
  {Kwiat}(2018)}]{kaneda18}%
  \BibitemOpen
  \bibfield  {author} {\bibinfo {author} {\bibfnamefont {F.}~\bibnamefont
  {Kaneda}}\ and\ \bibinfo {author} {\bibfnamefont {P.~G.}\ \bibnamefont
  {Kwiat}},\ }\bibfield  {title} {\enquote {\bibinfo {title} {High-efficiency
  single-photon generation via large-scale active time multiplexing},}\
  }\href@noop {} {\bibfield  {journal} {\bibinfo  {journal}
  {arXiv:quant-ph/1803.04803v1}\ } (\bibinfo {year} {2018})}\BibitemShut
  {NoStop}%
\bibitem [{\citenamefont {Pirandola}\ \emph {et~al.}(2015)\citenamefont
  {Pirandola}, \citenamefont {Eisert}, \citenamefont {Weedbrook}, \citenamefont
  {Furusawa},\ and\ \citenamefont {Braunstein}}]{pirandola15}%
  \BibitemOpen
  \bibfield  {author} {\bibinfo {author} {\bibfnamefont {S.}~\bibnamefont
  {Pirandola}}, \bibinfo {author} {\bibfnamefont {J.}~\bibnamefont {Eisert}},
  \bibinfo {author} {\bibfnamefont {C.}~\bibnamefont {Weedbrook}}, \bibinfo
  {author} {\bibfnamefont {A.}~\bibnamefont {Furusawa}}, \ and\ \bibinfo
  {author} {\bibfnamefont {S.}~\bibnamefont {Braunstein}},\ }\bibfield  {title}
  {\enquote {\bibinfo {title} {Advances in quantum teleportation},}\
  }\href@noop {} {\bibfield  {journal} {\bibinfo  {journal} {Nature Photonics}\
  }\textbf {\bibinfo {volume} {9}},\ \bibinfo {pages} {641} (\bibinfo {year}
  {2015})}\BibitemShut {NoStop}%
\bibitem [{\citenamefont {et~al.}(2019)}]{guo19}%
  \BibitemOpen
  \bibfield  {author} {\bibinfo {author} {\bibfnamefont {X.~Guo}\ \bibnamefont
  {et~al.}},\ }\bibfield  {title} {\enquote {\bibinfo {title} {Distributed
  quantum sensing in a continuous variable entangled network},}\ }\href@noop {}
  {\bibfield  {journal} {\bibinfo  {journal} {arXiv:quant-ph/1905.09408}\ }
  (\bibinfo {year} {2019})}\BibitemShut {NoStop}%
\bibitem [{\citenamefont {et~al.}(2017)}]{neuhaus17}%
  \BibitemOpen
  \bibfield  {author} {\bibinfo {author} {\bibfnamefont {L.~Neuhaus}\
  \bibnamefont {et~al.}},\ }\bibfield  {title} {\enquote {\bibinfo {title}
  {{PyRPL (Python Red Pitaya Lockbox) -- An open-source software package for
  FPGA-controlled quantum optics experiments}},}\ }\href@noop {} {\bibfield
  {journal} {\bibinfo  {journal} {CLEO/Europe-EQEC IEEE http://www.pyrpl.org}\
  } (\bibinfo {year} {2017})}\BibitemShut {NoStop}%
\bibitem [{\citenamefont {Mei}\ \emph {et~al.}(2007)\citenamefont {Mei},
  \citenamefont {Li}, \citenamefont {Huang},\ and\ \citenamefont
  {Rao}}]{mei07}%
  \BibitemOpen
  \bibfield  {author} {\bibinfo {author} {\bibfnamefont {H.}~\bibnamefont
  {Mei}}, \bibinfo {author} {\bibfnamefont {B.}~\bibnamefont {Li}}, \bibinfo
  {author} {\bibfnamefont {H.}~\bibnamefont {Huang}}, \ and\ \bibinfo {author}
  {\bibfnamefont {R.}~\bibnamefont {Rao}},\ }\bibfield  {title} {\enquote
  {\bibinfo {title} {Piezoelectric optical fiber stretcher for application in
  an atmospheric optical turbulence sensor},}\ }\href@noop {} {\bibfield
  {journal} {\bibinfo  {journal} {Appl. Opt.}\ }\textbf {\bibinfo {volume}
  {46}},\ \bibinfo {pages} {4371} (\bibinfo {year} {2007})}\BibitemShut
  {NoStop}%
\bibitem [{\citenamefont {Takei}\ \emph {et~al.}(2006)\citenamefont {Takei},
  \citenamefont {Lee}, \citenamefont {Moriyama}, \citenamefont
  {Neergaard-Nielsen},\ and\ \citenamefont {Furusawa}}]{takei06}%
  \BibitemOpen
  \bibfield  {author} {\bibinfo {author} {\bibfnamefont {N.}~\bibnamefont
  {Takei}}, \bibinfo {author} {\bibfnamefont {N.}~\bibnamefont {Lee}}, \bibinfo
  {author} {\bibfnamefont {D.}~\bibnamefont {Moriyama}}, \bibinfo {author}
  {\bibfnamefont {J.~S.}\ \bibnamefont {Neergaard-Nielsen}}, \ and\ \bibinfo
  {author} {\bibfnamefont {A.}~\bibnamefont {Furusawa}},\ }\bibfield  {title}
  {\enquote {\bibinfo {title} {Time-gated {Einstein-Podolsky-Rosen}
  correlation},}\ }\href@noop {} {\bibfield  {journal} {\bibinfo  {journal}
  {Phys. Rev. A}\ }\textbf {\bibinfo {volume} {74}},\ \bibinfo {pages} {060101}
  (\bibinfo {year} {2006})}\BibitemShut {NoStop}%
\end{thebibliography}
\end{document}


\title{Supplementary information for\\Fiber coupled EPR-state generation using a single temporally multiplexed squeezed light source}
	\author{Mikkel V. Larsen, Xueshi Guo, Casper R. Breum, Jonas S. Neergaard-Nielsen\\and Ulrik L. Andersen}
	\date{June 6, 2019}
	\maketitle
	\vspace{1cm}
	\tableofcontents	
	
	\newpage
\section{Quadrature relations}
\begin{figure}[b!]
	\centering
	\begin{subfigure}{0.40\textwidth}
		\centering
		\includegraphics[width=\textwidth]{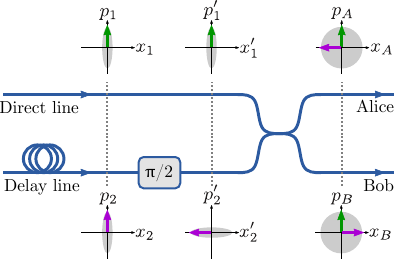}
		\caption{\label{fig:Qtrans}}
	\end{subfigure}
	\quad\quad\quad\quad
	\begin{subfigure}{0.35\textwidth}
		\centering
		\includegraphics[width=\textwidth]{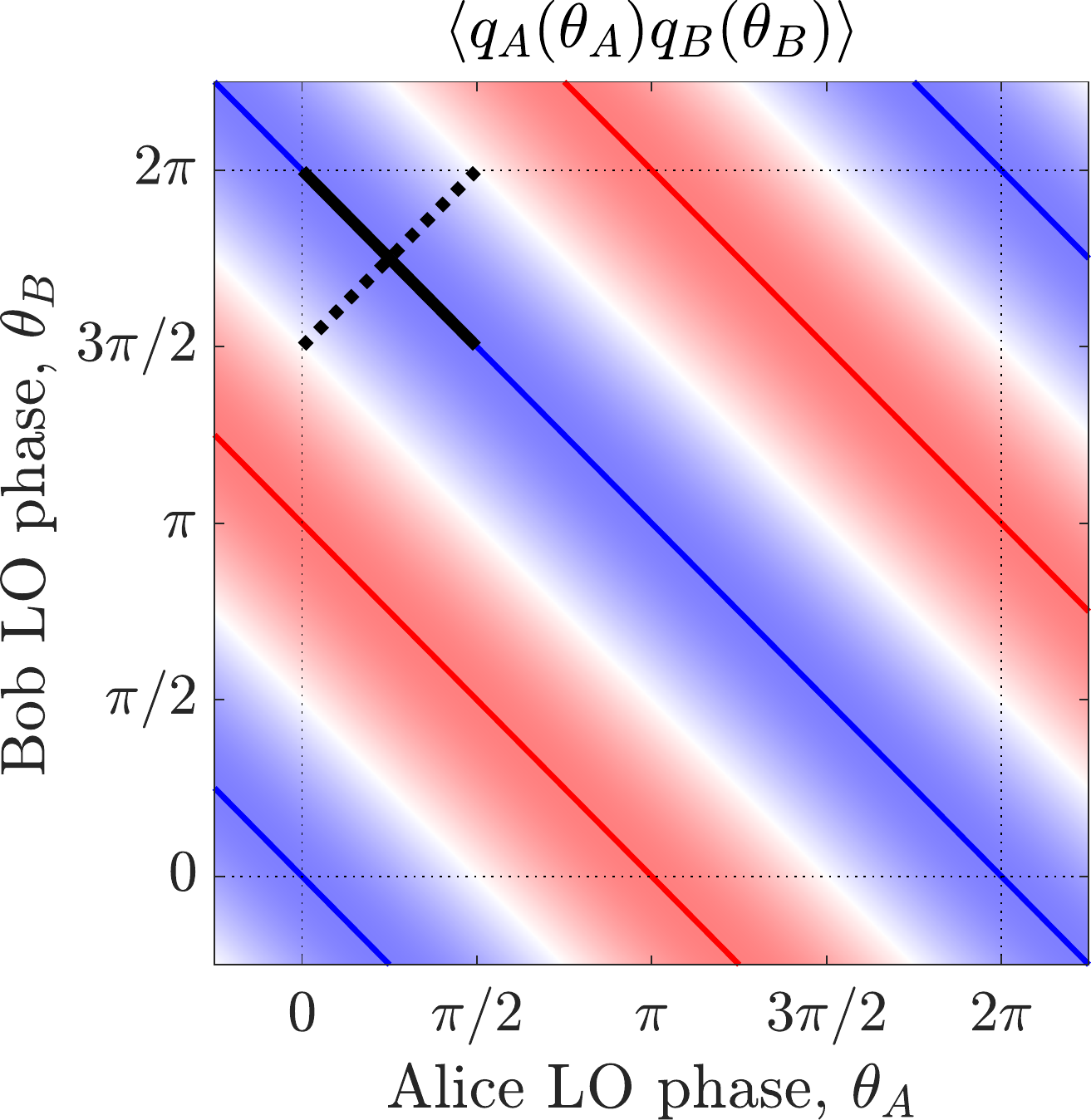}
		\caption{\label{fig:corr}}
	\end{subfigure}
	\caption{(a) Quadrature transformations of the two-mode squeezed state generation, where simultaneous amplitude squeezed states in the direct and delay line are interfered at a symmetric beam splitter (fiber coupler) with a $\pi/2$ phase difference in the delay line. (b) Quadrature covariance in eq. (\ref{eq:cov}) with red and blue indicating areas of correlation and anti-correlation respectively, and solid coloured lines indicates maximum correlations. The two measured sets of quadratures for partial tomography are marked by the black solid and dotted line for set 1 ($\theta_B=-\theta_A$) and set 2 ($\theta_B=\theta_B-\pi/2$) respectively.}
\end{figure}

To generate two-mode squeezed states from a single squeezer, two amplitude squeezed states from the same squeezing source, but in different temporal modes, are guided into two different spatial modes, synchronized in time by a delay in one spatial mode, and finally interfered on a symmetric beam splitter with a $\pi/2$ phase difference. The phase-space in the two spatial modes after the temporal modes of squeezing are synchronized, after $\pi/2$ rotation, and after interfering the two modes, are illustrated in Figure \ref{fig:Qtrans}. In the following quadrature relations of the generated two-mode squeezed states are derived, assuming the input states being squeezed vacuum, perfect phase control, and the beam splitter (or fiber coupler) being  symmetric.

After synchronization, the two spatial modes -- the direct ($x_1,p_1$) and delay ($x_2,p_2$) line -- are both amplitude squeezed states, i.e.
\begin{equation*}
	\braket{\Delta x_1^2},\braket{\Delta x_2^2}<V_0\quad,\quad\braket{\Delta p_1^2},\braket{\Delta p_2^2}>V_0\;,
\end{equation*}
where $V_0$ is the vacuum variance. After a $\pi/2$ phase-space rotation in the delay line, the new quadratures read
\begin{equation*}\begin{aligned}
	x_1'&=x_1\quad &,\quad p_1'&=p_1\;,\\
	x_2'&=-p_2\quad &,\quad p_2'&=x_2\;.
\end{aligned}\end{equation*}
Finally, after a beam splitter transformation of the symmetric fiber coupler, we get
\begin{equation}\begin{aligned}\label{eq:all_quad}
	&x_A=\frac{1}{\sqrt{2}}\left(x_1'+x_2'\right)=\frac{1}{\sqrt{2}}\left(x_1-p_2\right)\quad &,\quad &p_A=\frac{1}{\sqrt{2}}\left(p_1'+p_2'\right)=\frac{1}{\sqrt{2}}\left(p_1+x_2\right)\;,\\
	&x_B=\frac{1}{\sqrt{2}}\left(x_1'-x_2'\right)=\frac{1}{\sqrt{2}}\left(x_1+p_2\right)\quad &,\quad &p_B=\frac{1}{\sqrt{2}}\left(p_1'-p_2'\right)=\frac{1}{\sqrt{2}}\left(p_1-x_2\right)\;.
\end{aligned}\end{equation}
Thus, when measuring the noise of $x_A+x_B=\sqrt{2}x_1$ and $p_A-p_B=\sqrt{2}x_2$, the squeezing of $x_1$ and $x_2$ are visible, while when measuring the noise of $x_A-x_B=-\sqrt{2}p_2$ and $p_A+p_B=\sqrt{2}p_1$, the anti-squeezing of $p_1$ and $p_2$ are visible. Or in other words: $x_A$ and $x_B$ are correlated while $p_A$ and $p_B$ are anti-correlated, both below the standard quantum limit demonstrating quadrature entanglement.

The two modes of the two-mode squeezed state at the fiber coupler output are sent to two homodyne detection stations, Alice and Bob, where an arbitrary quadrature, $q_i(\theta_i)$, is measured depending on the local oscillator phase $\theta_i$ at Alice ($i=A$) and Bob ($i=B$),
\begin{equation}\label{eq:qA}
	q_A(\theta_A)=x_A\cos\theta_A+p_A\sin\theta_A=\frac{1}{\sqrt{2}}\left(x_1-p_2\right)\cos\theta_A+\frac{1}{\sqrt{2}}\left(p_1+x_2\right)\sin\theta_A\;,
\end{equation}
\begin{equation}\label{eq:qB}
	q_B(\theta_B)=x_B\cos\theta_B+p_B\sin\theta_B=\frac{1}{\sqrt{2}}\left(x_1+p_2\right)\cos\theta_B+\frac{1}{\sqrt{2}}\left(p_1-x_2\right)\sin\theta_B\;.
\end{equation}
The quadrature entanglement can then be illustrated as correlations, or covariance, between $q_A(\theta_A)$ and $q_B(\theta_B$):
\begin{equation*}\begin{aligned}
	\text{cov}\left\lbrace q_A(\theta_A),q_B(\theta_B)\right\rbrace &= \braket{q_A(\theta_A)q_B(\theta_B)}-\braket{q_A(\theta_A)}\braket{q_B(\theta_B)}=\braket{q_A(\theta_A)q_B(\theta_B)}\\
	&=\frac{1}{2}\Big\langle\left[\left(x_1-p_2\right)\cos\theta_A+\left(p_1+x_2\right)\sin\theta_A\right]\left[\left(x_1+p_2\right)\cos\theta_B+\left(p_1-x_2\right)\sin\theta_B\right]\Big\rangle\\
	&=\frac{1}{2}\left(\braket{x_1^2}-\braket{p_2^2}\right)\cos\theta_A\cos\theta_B+\frac{1}{2}\left(\braket{P_1^2}-\braket{x_2^2}\right)\sin\theta_A\sin\theta_B\\
	&=\frac{1}{2}\left(\braket{\Delta x_1^2}-\braket{\Delta p_2^2}\right)\cos\theta_A\cos\theta_B-\frac{1}{2}\left(\braket{\Delta x_2^2}-\braket{\Delta p_1^2}\right)\sin\theta_A\sin\theta_B\;,
\end{aligned}\end{equation*}
where it is used that $\braket{q_A(\theta_A)}=\braket{q_B(\theta_B)}=0$ for two-mode squeezed states, $\braket{q_1q_2}=0$ for $q_1=(x_1,p_1)$ and $q_2=(x_2,p_2)$ since the amplitude squeezed states before interfering are separable squeezed vacuum states, $\braket{x_ip_i}=0$ for $i=(1,2)$ as amplitude and phase quadrature are independent for squeezing exactly along $x$ or $p$, and $\braket{\Delta q_i^2}=\braket{q_i^2}-\braket{q_i}^2=\braket{q_i^2}$ for $q=(x,p)$ and $i=(1,2)$ as $\braket{q_i}=0$ for squeezed vacuum. Thus, since $\braket{\Delta p_1^2},\braket{\Delta p_1^2}>\braket{\Delta x_1^2},\braket{\Delta x_1^2}$, we expect strong anti-correlation when both $\theta_A$ and $\theta_B$ equals $\pi n$ with $n$ being an integer number, while strong correlation when both $\theta_A$ and $\theta_B$ equals $\pi(n+1/2)$. This is easier seen in the ideal case where the direct and delay line are identical so that $\braket{\Delta x_1^2}=\braket{\Delta x_2^2}\equiv V_x$ and $\braket{\Delta p_1^2}=\braket{\Delta p_2^2}\equiv V_p$, in which case the above covariance can be simplified using $\cos(\alpha+\beta)=\cos\alpha\cos\beta-\sin\alpha\sin\beta$:
\begin{equation}\label{eq:cov}
	\text{cov}\left\lbrace q_A(\theta_A),q_B(\theta_B)\right\rbrace=\braket{q_A(\theta_A)q_B(\theta_B)}=\frac{1}{2}\left(V_x-V_p\right)\cos(\theta_A+\theta_B)\;.
\end{equation}

The quadrature covariance for the ideal case in (\ref{eq:cov}) is shown in Figure \ref{fig:corr}. Here, two-mode squeezing and anti-squeezing can be measured in areas of strong correlations, and is constant along constant $\theta_A+\theta_B$. As a result, for partial tomography of the generated two-mode squeezed state, we measure two sets of quadratures $(q_A(\theta_A),q_B(\theta_B))$: Set 1 along constant correlation (maximum squeezing and anti-squeezing) with $\theta_A=-\theta_B=\theta$; set 2 perpendicular to constant squeezing (across maximum squeezing and anti-squeezing) with $\theta_A=\theta_B+\pi/2=\theta$. Both set are measured with $\theta$ in the range from $0$ to $\pi/2$, and is marked in Figure \ref{fig:corr}. By adding and subtracting $q_A$ and $q_B$ of these two measured sets, we observe two-mode squeezing, anti-squeezing and all in between.

When adding/subtracting quadratures from Alice and Bob measured in set 1 ($\theta_B=-\theta_A$), from eq. (\ref{eq:qA}-\ref{eq:qB}) the added and subtracted quadrature noise becomes
\begin{equation}\label{eq:set1p}
	\braket{\Delta (q_A(\theta)+q_B(-\theta))^2}=2\left(\braket{\Delta x_1^2}\cos^2\theta+\braket{\Delta x_2^2}\sin^2\theta\right)\;,
\end{equation}
\begin{equation}\label{eq:set1n}
	\braket{\Delta (q_A(\theta)-q_B(-\theta))^2}=2\left(\braket{\Delta p_2^2}\cos^2\theta+\braket{\Delta p_1^2}\sin^2\theta\right)\;,
\end{equation}
and thus we expect to measure two-mode squeezing and anti-squeezing for all $\theta$, and constant squeezing levels in the case of symmetric two-mode squeezing where $\braket{\Delta x_1^2}=\braket{\Delta x_2^2}$ and $\braket{\Delta p_1^2}=\braket{\Delta p_2^2}$.

Finally, when adding/subtracting quadratures from Alice and Bob measured in set 2 ($\theta_B=\theta_A-\pi/2$) we obtain
\begin{equation*}\begin{aligned}
	q_A(\theta)\pm q_B(\theta-\pi/2)&=x_A\cos\theta+p_A\sin\theta\pm x_B\cos(\theta-\pi/2)\pm p_B\sin(\theta-\pi/2)\\
	&=x_A\cos\theta+p_A\sin\theta\pm x_B\sin\theta\mp p_B\cos\theta\;.
\end{aligned}\end{equation*}
For $\theta=\pi/4$ we extract again two-mode squeezing and anti-squeezing:
\begin{equation*}
	\braket{\Delta q_A(\pi/4)+q_B(\pi/4-\pi/2))^2} = \frac{1}{2}\braket{\Delta ( x_A+p_A+x_B-p_B)^2}=2\left(\braket{\Delta x_1^2}+\braket{\Delta x_2^2}\right)\;,
\end{equation*}
\begin{equation*}
	\braket{\Delta q_A(\pi/4)-q_B(\pi/4-\pi/2))^2} = \frac{1}{2}\braket{\Delta ( x_A+p_A-x_B+p_B)^2}=2\left(\braket{\Delta p_1^2}+\braket{\Delta p_2^2}\right)\;,
\end{equation*}
using (\ref{eq:all_quad}). Moving away from $\theta=\pi/4$ we measure less and less two-mode squeezing and anti-squeezing, until $\theta=0$ or $\pi/2$ where, using (\ref{eq:all_quad}),
\begin{equation*}\begin{aligned}
	\braket{\Delta(q_A(0)\pm q_B(0-\pi/2))^2}&=\braket{\Delta(x_A\mp p_B)^2})=\frac{1}{2}\left(\braket{\Delta x_1^2}+\braket{\Delta p_2^2}+\braket{\Delta p_1^2}+\braket{\Delta x_2^2}\right)\;,\\
	\braket{\Delta(q_A(\pi/2)\pm q_B(\pi/2-\pi/2))^2}&=\braket{\Delta(p_A\pm x_B)^2})=\frac{1}{2}\left(\braket{\Delta p_1^2}+\braket{\Delta x_2^2}+\braket{\Delta x_1^2}+\braket{\Delta p_2^2}\right)\;,
\end{aligned}\end{equation*}
and we simply measure added uncorrelated noise at Alice and Bob corresponding to thermal states when tracing out one mode of the two-mode squeezed state.

\section{Experimental methods}\label{sec:expMethods}
\begin{figure}
	\centering
	\begin{subfigure}{0.6\textwidth}
		\includegraphics[width=\textwidth]{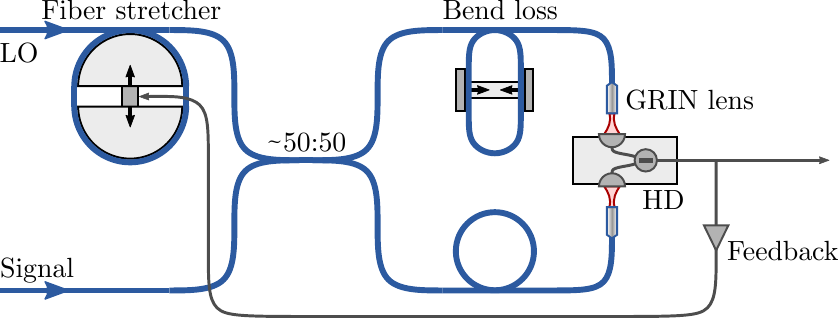}
		\caption{}
		\label{fig:HD}
	\end{subfigure}
	\quad\quad
	\begin{subfigure}{0.2\textwidth}
		\includegraphics[width=\textwidth]{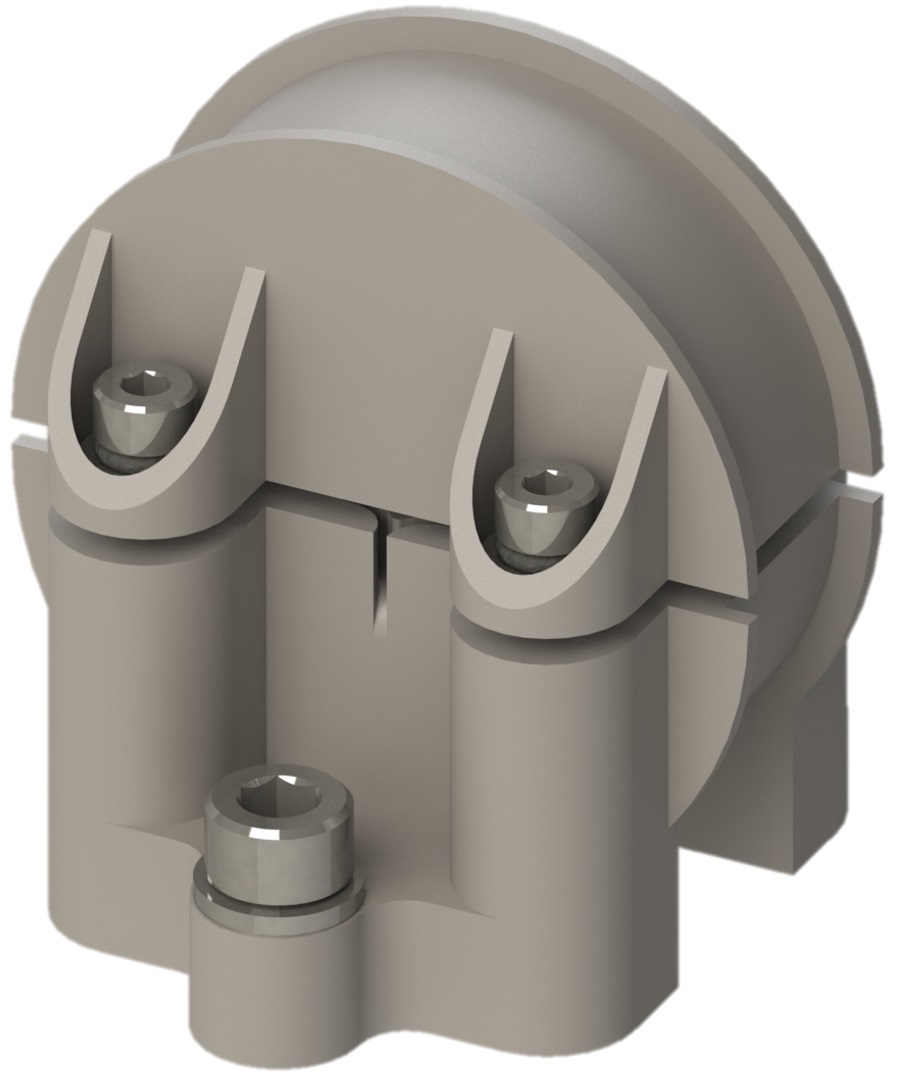}
		\caption{}
		\label{fig:FS}
	\end{subfigure}
	\caption{(a) Schematics of the fiber coupled homodyne detection (HD) setup, where blue lines indicates single mode fiber. (b) Illustration of the homemade fiber stretcher consisting of a top and bottom part around a piezoelectric actuator. Pre-load can be applied by tightening the screw with O-rings attaching the top part to the bottom part.}
\end{figure}

The experimental setup is outlined in the main text. In Fig. \ref{fig:HD} schematics of the fiber-coupled homodyne detection (HD) is seen. Here, the local oscillator (LO) phase is controlled by a fiber stretcher illustrated in Fig. \ref{fig:FS}, and mixed with the signal in a $\sim$50:50 fiber coupler. The HD is balanced by inducing bend loss in one fiber coupler output arm: The fiber is coiled up in between two plates, which are squeezed towards each other with a micrometer screw. Finally, light is coupled from fiber onto the photo diodes using anti-reflective coated graded-index (GRIN) lenses.

The fiber stretcher based on \cite{mei07} is illustrated in Fig. \ref{fig:FS}. With a piezoelectric actuator the fiber stretcher top and bottom part are pushed away from each other, stretching a single mode fiber coiled around it. The bandwidth is set by the mechanical resonance frequency of the top part and optimized to $\SI{2.5}{kHz}$ by applying pre-load. This can be improved by a lighter design of the top-part, but it is sufficient for our purpose.

\subsection{Temporal data processing}
As two-mode squeezing is generated by temporal multiplexing a single squeezing source, we need to filter temporal modes for analysis. Due to the switching and synchronization of temporal modes, two-mode squeezing at the 50:50 fiber coupler output is only present half of the time in a switching period (whereas vacuum is present the other half time, which is the cost of using a single squeezing source).

\begin{figure}[t!]
	\centering
	\includegraphics[width=\textwidth]{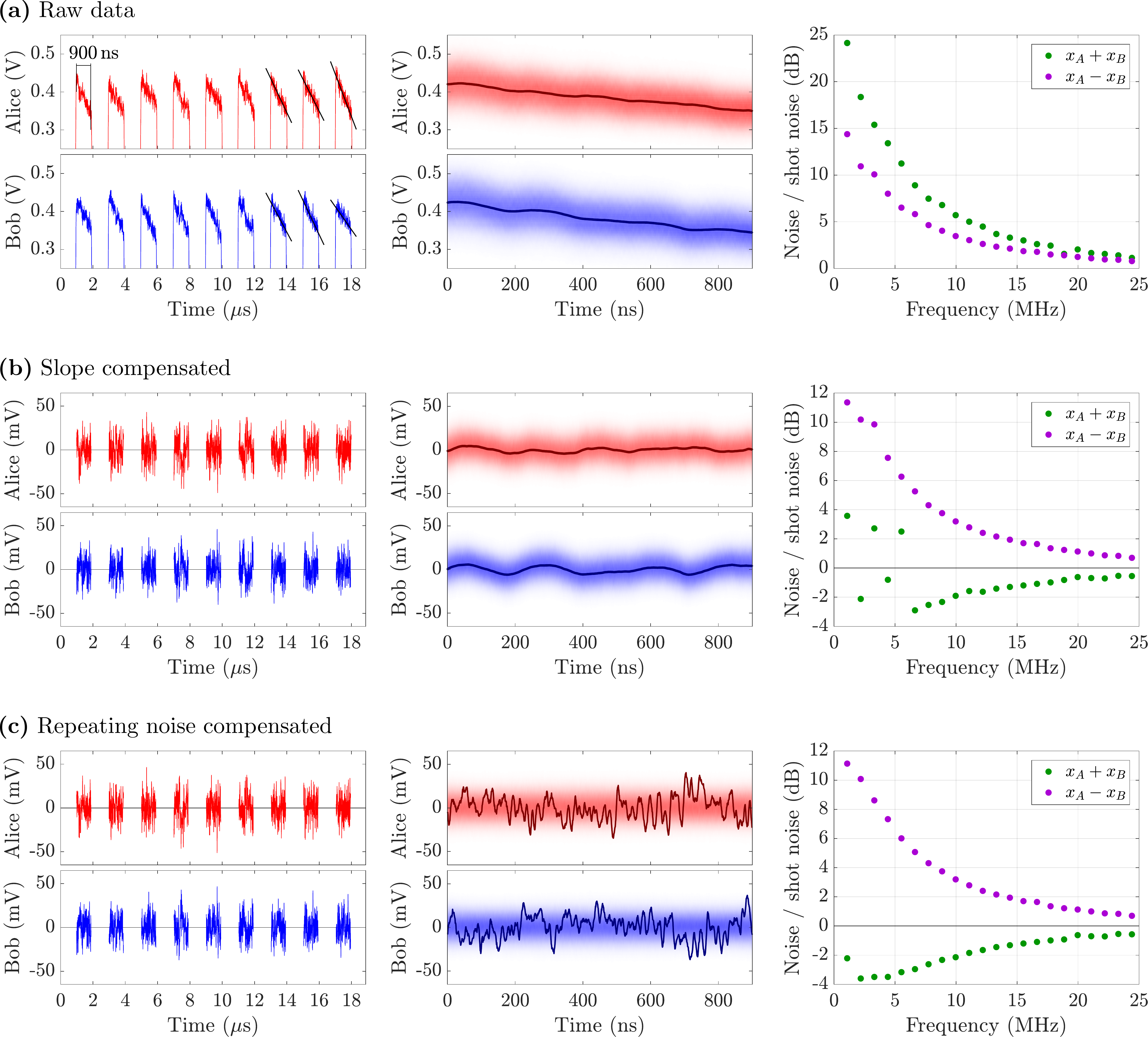}
	\caption{\label{fig:dataProc} Example of the temporal post processing procedure of measured data, here of $x_A$ and $x_B$. (a) Raw data with HD signals from Alice and Bob shown in (left), temporal histogram of $\SI{16000}{}$ time traces of $\SI{900}{ns}$ length as indicated in (left) shown in (center), and spectrum of added and subtracted signals in (right). (b) Data in (a) compensated for non-zero slope by subtracting linear regression lines as shown in (a,left). (c) Data in (b) compensated for repeating noise by subtracting the average of the $\SI{16000}{}$ slope compensated time traces shown as solid lines in (b,center). Here, the solid lines in (center) is a single time trace from Alice and Bob indicating quadrature correlations, while the final two-mode squeezing spectrum after processing is shown in (right).}
\end{figure}

The HD signal from Alice and Bob is shown in Figure \ref{fig:dataProc}a (left) as function of time, where the two-mode squeezing is seen with an offset due to the coherent seed beam transmitted through the setup for phase locks. Here we can select time traces of $\SI{900}{ns}$ well within the temporal region of two-mode squeezing with $\SI{1}{\micro s}$ length. Doing so, one measurement set consist of such $\SI{16000}{}$ time traces, with the temporal histogram shown in Fig. \ref{fig:dataProc}a (center). The quadratures measured in Figure \ref{fig:dataProc} are $x_A$ and $x_B$, and thus by adding and subtracting the measured quadrature traces, we expect squeezing and anti-squeezing according to eq. (\ref{eq:set1p}-\ref{eq:set1n}). However, as seen in Figure \ref{fig:dataProc}a (right) where the Fourier transform of $x_A\pm x_B$ is shown, this is not the case. This is due to the slope seen on each time trace in Figure \ref{fig:dataProc}a (left) and (center), which is caused by a decaying detector response when the coherent seed beam is turned on and off in the detector.

To compensate for the non-zero slope on each time trace, we cannot simply high pass filter the HD signals, as the slope includes frequency components in the MHz-regime, and we would filter out the two-mode squeezing. Instead, we subtract the slope from each individual time trace. However, due to spurious interference of the coherent seed beam and small phase fluctuations, the slope of each time trace varies, and we cannot subtract the same slope from each time trace. Instead, assuming the decaying detector response is linear on this time scale, a linear regression line is fitted to each time trace, as shown on the last three pulses in Figure \ref{fig:dataProc}a (left), and subtracted.

The result of compensating for non-zero slope on each time trace is shown in Figure \ref{fig:dataProc}b, where the HD signals in (left) are effectively high pass filtered without affecting the squeezing spectrum, and two-mode squeezing below added shot noise is visible in (right). Yet, peaks are seen in the spectrum due to oscillations visible in the temporal histogram in Figure \ref{fig:dataProc}b (center). These are caused by the switching process and maybe oscillating detection response. However, since they are repeating with the switching process, they can be compensated for by subtracting an average of the $\SI{16000}{}$ time traces, indicated by the solid lines in Figure \ref{fig:dataProc}b (center).

Finally, the result of compensating for the repeating noise is shown in Figure \ref{fig:dataProc}c, where in (center) the temporal histogram is seen to be nicely constant, and a two-mode squeezing spectrum in (right) resembling the OPO spectrum without additional noise peaks. Here the degrading of squeezing in the first frequency component shown is due to the coherent seed beam not being shot noise limited, and limiting bandwidth of $\sim\SI{1}{MHz}$ due to the $\SI{900}{ns}$ short time traces. The solid line shown in Figure \ref{fig:dataProc}c (center) is a single time trace at Alice and Bob showing quadrature correlations. To observe stronger correlations, only frequency components within the OPO bandwidth should be observed, which is the case in the main text Figure \tomography.

\section{Sources of noise and theoretical prediction}
\begin{figure}
	\centering
	\begin{subfigure}{0.4\textwidth}
		\centering
		\includegraphics[width=\textwidth]{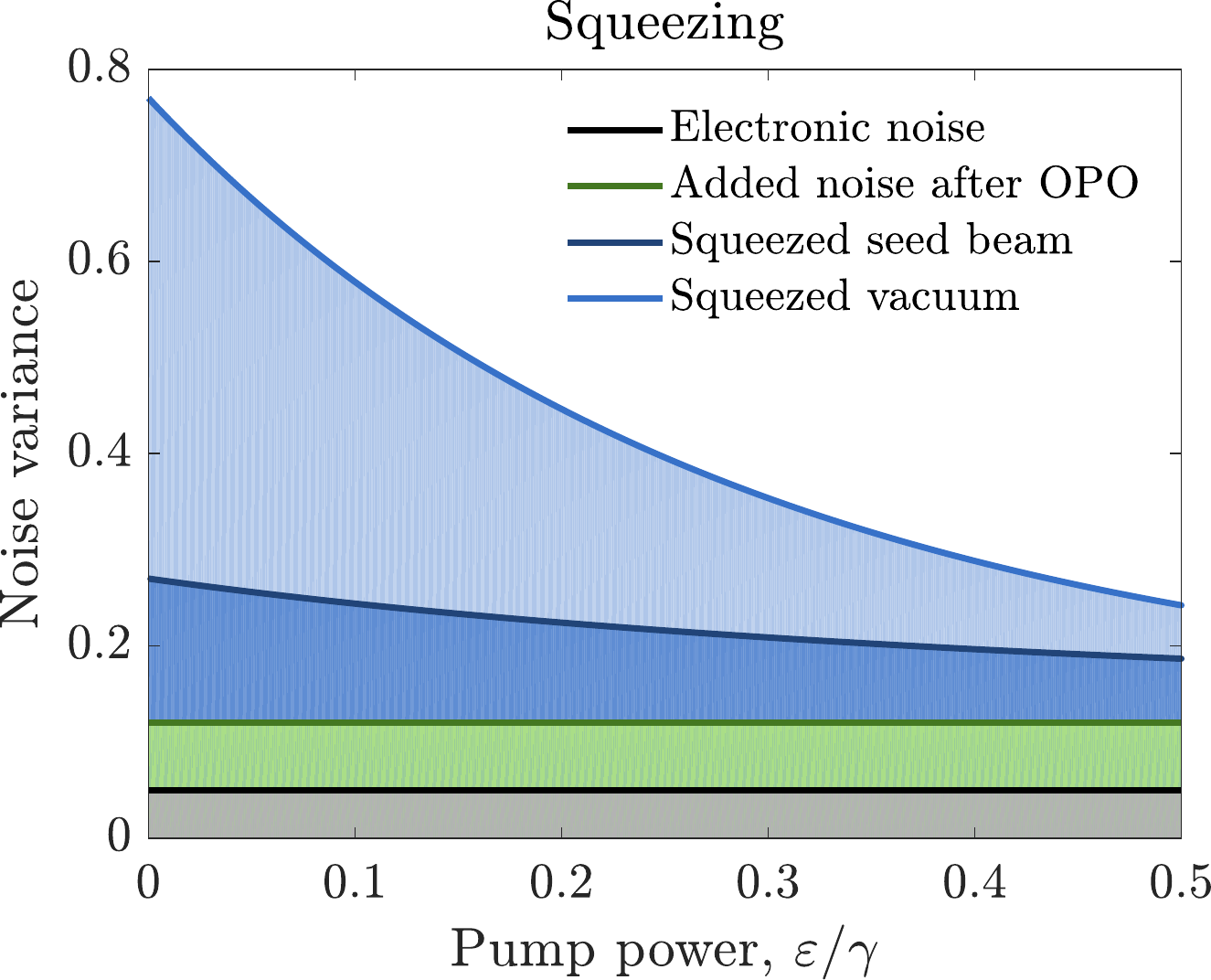}
		\caption{\label{fig:sqNoise}}
	\end{subfigure}
	\quad\quad
	\begin{subfigure}{0.4\textwidth}
		\centering
		\includegraphics[width=\textwidth]{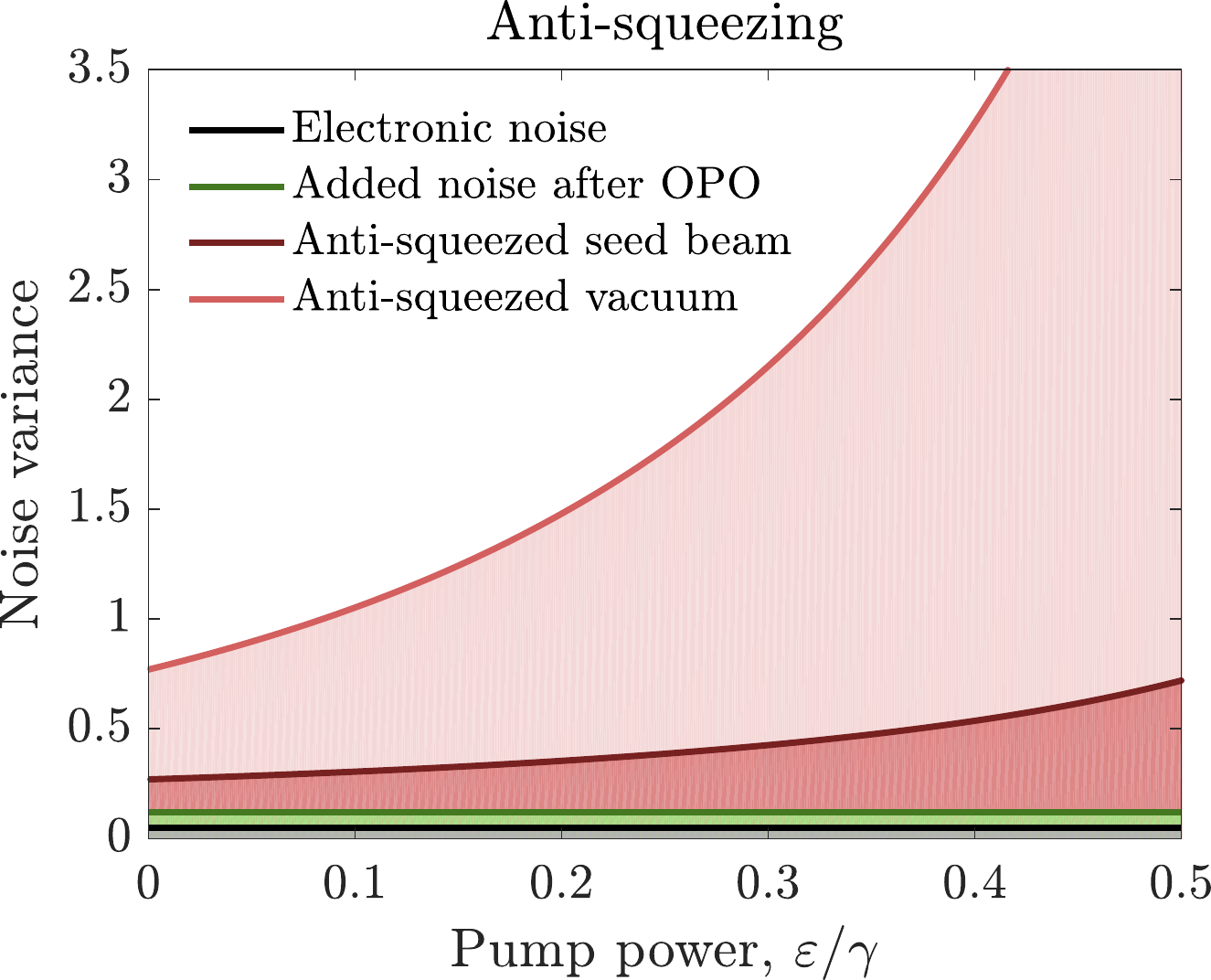}
		\caption{\label{fig:asqNoise}}
	\end{subfigure}
	\caption{\label{fig:Noise}Illustration of quadrature noise when measuring single mode squeezing in the direct or delay line, including vacuum squeezing, squeezing of seed beam noise, added noise after the OPO and electronic noise. (a) Illustrates how the different noise sources behave in the squeezed quadrature when increasing the pump power, and in (b) the anti-squeezed quadrature is illustrated. The sources or noise shown here are not to scale.}
\end{figure}

To theoretically predict the measured two-mode squeezing, we include several sources of noise. Besides the squeezed vacuum noise itself, the seed beam transmitted through the optical parametric oscillator (OPO) contains noise, while additional noise is added after the OPO. Besides this, due to phase fluctuations, when measuring one quadrature, noise of the other quadrature is observed as well. Finally, electronic noise is also considered. When adding and subtracting the measured $x$- or $p$-quadrature of the two-mode squeezed state at Alice and Bob, the single mode squeezed states in the direct or delay line is extracted according to (\ref{eq:set1p}-\ref{eq:set1n}) for $\theta=0$ and $\pi/2$. So, for simplicity, in the following discussion we consider the case of measuring single mode squeezing in the direct and delay line. The  different sources of quadrature noise are summarized in Figure \ref{fig:Noise} as a function of the pump power.

Electronic noise is present in both measured quadratures and shot noise. It is independent on pump power, and we deal with it by subtracting the electronic noise variance from both measured quadrature variance and shot noise variance. However, the electronic noise is less than $\SI{-20}{dB}$ relative to shot noise, and can in principle be neglected.

Noise unrelated to the squeezing process is characterized by measuring the seed spectrum by blocking the pump to the OPO, and the result is seen in the main text Figure \spectrum { }(grey points). From the direct line, low frequency noise resulting from technical noise of the seed beam is observed. Even more low frequency noise is apparent in the squeezed state of the delay line, and we believe it originates from phase noise generated by the $\SI{200}{m}$ fiber, and amplitude noise from the fiber switch which is most prominent at $5\text{-}\SI{6}{MHz}$.

The measured seed beam noise undergoes squeezing in the OPO, and thus depends on the pump power. The squeezed and anti-squeezed quadrature noise spectrum from the OPO including the noisy seed beam is derived below in section \ref{sec:sq_spectrum} to be
\begin{equation}\label{eq:sqSpectrum}
	\braket{\Delta q^2}=\frac{1}{2}\mp\frac{2\varepsilon\gamma\eta}{(\gamma\pm\varepsilon)^2+\omega^2}+\frac{K_q}{(\gamma\pm\varepsilon)^2+\omega^2}\quad,\quad q=x,p,
\end{equation}
where $\varepsilon$ is the pump rate, $\gamma$ is the total OPO decay rate, $\eta$ is the overall efficiency and $\omega$ is the angular frequency. The first two terms corresponds to vacuum squeezing when subtracted and anti-squeezing when added (with $1/2$ being the vacuum variance), while the third term corresponds to squeezing/anti-squeezing of the seed beam noise $\braket{\Delta q_s^2}$ with $K_q=4\eta\gamma\gamma_s(\braket{\Delta q_s^2}-1/2)$, where $\gamma_s$ is the coupling rate of the mirror through which the seed beam is leaked into the OPO. From the measured seed beam spectrum when blocking the pump ($\varepsilon=0$), $K_q$ can be estimated as
\begin{equation*}
	K_q=(\gamma^2+\omega^2)(\braket{\Delta q_0^2}-1/2)\;,
\end{equation*}
where $\braket{\Delta q_0^2}$ then corresponds to grey points in the direct line of the main text Figure \spectrum . Notice how the squeezed vacuum in (\ref{eq:sqSpectrum}) (the first two terms) goes towards 0 for $\varepsilon\rightarrow\gamma$ and $\eta=1$ at $\omega=0$, while the seed beam noise goes towards $(\braket{\Delta q_0^2}-1/2)/4$, and thus the seed beam noise can only be eliminated when $\braket{\Delta q_0^2}=1/2$ is shot noise limited. The OPO decay rate is estimated to $\gamma=2\pi\SI{8.1}{MHz}$ by measuring the OPO intracavity losses (0.55\%), the cavity length ($\SI{320}{mm}$) and the transmittivity of the coupling mirror (10\%), while we estimate the pump rate to $\varepsilon=2\pi\SI{5.2}{MHz}$ for a pump power of $\SI{350}{mW}$ and an estimated OPO threshold power of $\SI{833}{mW}$. The overall efficiency is $\eta=68\%$ as discussed in section \ref{sec:expMethods}.

To predict the squeezing, the direct line eq. (\ref{eq:sqSpectrum}) is sufficient, since negligible noise is added after the OPO. However, as apparent in the main text Figure \spectrum { }and discussed above, both amplitude and phase noise is added in the delay line, and is independent on the pump power. To encounter this, the difference in measured noise spectrum in the direct and delay line when blocking the pump (grey points in main text Figure \spectrum ) is added to eq. (\ref{eq:sqSpectrum}) for the squeezing spectrum in the delay line.

Finally, phase fluctuations are included by applying a normal distribution $P(\sigma,\theta)$ of width $\sigma$ to the measured  quadrature as
\begin{equation*}\begin{aligned}
	\braket{\Delta x_\sigma^2(\theta)}&=\int P(\sigma,\theta)\left(\braket{\Delta x^2}\cos^2\theta+\braket{\Delta p^2}\sin^2\theta\right)\text{d}\theta\\
	&\approx\braket{\Delta x^2}\cos^2(\theta+\sigma)+\braket{\Delta p^2}\sin^2(\theta+\sigma)\;,
\end{aligned}\end{equation*}
and similar for $\braket{\Delta p^2}$ with $\cos(\theta+\sigma)$ and $\sin(\theta+\sigma)$ interchanged, and with the the approximation valid for small $\sigma$ \cite{aoki06}. Since we expect different phase fluctuations in the direct line and the $\SI{200}{m}$ long delay line, different normal distributions of width $\sigma_1$ and $\sigma_2$ are used respectively. By using $\sigma_1$ and $\sigma_2$ as fitting parameters, the theoretical predicted squeezing in eq. (\ref{eq:sqSpectrum}), with additional added noise in the delay line, is fitted to the experimental data, and shown as hollow points in the main text Figure \spectrum . From the fitting we get $\sigma_1=1.9\pm1.2^\circ$ and $\sigma_2=4.1\pm0.6^\circ$ (uncertainty estimated as $95\%$ confidence interval), and the theoretical prediction is seen to agree excellent with measured data. 

\subsection{Squeezing spectrum with seed beam noise}\label{sec:sq_spectrum}
\begin{figure}[b!]
	\centering
	\includegraphics[width=6.5cm]{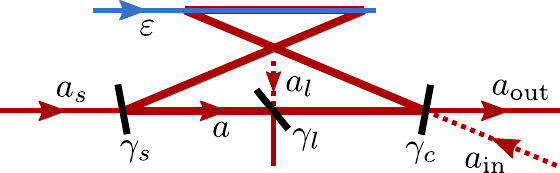}
	\caption{\label{fig:OPO} Schematic model of the OPO as squeezing source. Here, $\gamma_c$ is the decay rate due to the cavity coupling mirror of 10\% transmission, while $\gamma_s$ is the decay rate due to the high reflective mirror through which the seed is leaked into the cavity. The internal cavity loss is modelled by a beams splitter transformation leading to the decay rate $\gamma_l$. $a$ is the annihilation operator of  the cavity mode, $a_i$  for $i=\text{in},\text{out},s,l$ are annihilation operators of modes mixing with the cavity mode, and $\varepsilon$ is the effective pump intensity which is treated classically. Dotted line corresponds to modes with vacuum.}
\end{figure}

In order to derive the squeezing spectrum of the OPO in eq. (\ref{eq:sqSpectrum}) with a seed beam leaked into the cavity, we follow the approach in \cite{collett84} where the spectrum is derived from the quantum Langevin equations of a cavity. However, in \cite{collett84} the leakage of a seed beam into the cavity is not considered, and so the model is here extended for this purpose -- a schematic of the model is sketched in Fig. \ref{fig:OPO}.

Treating the pump power classically with the effective pump intensity $\varepsilon$, and frequency of double the cavity resonant mode frequency $\omega_0$, the system Hamiltonian of the OPO cavity mode is
\begin{equation*}
	H_\text{sys}=\hbar\omega_0a^\dagger a+\frac{1}{2}\left(\varepsilon e^{-i2\omega_0t}a^{\dagger2}-\varepsilon^*e^{i2\omega_0t}a^2\right)\;,
\end{equation*}
where $a$ is the annihilation operator of the cavity mode. Here the second term is the Hamiltonian of the parametric down conversion process of the pump in a second order non-linear crystal, and the pump is assumed non-depleted (i.e. far below the OPO threshold). Using $[a,H_\text{sys}]=\hbar\omega_0a+i\hbar\varepsilon e^{-i2\omega_0t}a^\dagger$, the Langevin equations becomes
\begin{equation}\begin{aligned}\label{eq:Langevin}
	\frac{da}{dt}&=-\frac{i}{\hbar}[a,H_\text{sys}]-\gamma a +\Gamma\\
	&=-i\omega_0a+\varepsilon e^{-i2\omega_0t}a^\dagger -\gamma a+\Gamma\;,
\end{aligned}\end{equation}
where $\gamma$ is the cavity damping rate, and $\Gamma$ is the noise operator. Cavity damping is caused by out coupling of the cavity mode, internal cavity loss, and leakage through the high reflective mirror through which the seed is coupled, each leading to corresponding decay rates $\gamma_c$, $\gamma_l$ and $\gamma_s$ respectively so that $\gamma=\gamma_c+\gamma_l+\gamma_s$. Where light is coupled out, vacuum in the mode $a_\text{in}$ is coupled in. Similar, the cavity internal loss can be modelled by a beam splitter transformation coupling vacuum in mode $a_l$ into the cavity mode. Finally, with the seed beam in mode $a_s$ leaked through a high reflector, the noise operator is $\Gamma=\sqrt{2\gamma_c}a_\text{in}+\sqrt{2\gamma_l}a_l+\sqrt{2\gamma_s}a_s$. With the mirror between $a_s$ and $a$ being high reflective, we have $\gamma_s\ll\gamma_c,\gamma_l$ and so $\gamma\approx\gamma_c+\gamma_l$. However, we cannot choose $\gamma_s=0$ so that $\Gamma=\sqrt{2\gamma_c}a_\text{in}+\sqrt{2\gamma_l}a_l$, as then no seed beam leaks into the cavity mode and the model simplifies to that in \cite{collett84}.

Moving to a rotating frame with frequency $\omega_0$,
\begin{equation*}
	a\rightarrow ae^{-i\omega_0t}\quad,\quad\frac{da}{dt}\rightarrow \frac{d}{dt}\left\lbrace ae^{-i\omega_0t}\right\rbrace=\frac{da}{dt}e^{-i\omega_0t}-i\omega_0ae^{-i\omega_0t}\;,
\end{equation*}
eq. (\ref{eq:Langevin}) simplifies to
\begin{equation}\label{eq:MLangevin}
	\frac{d\bm{a}}{dt}=\left( \textbf{A}-\gamma\bm{1}\right)\bm{a}+\Gamma\;,
\end{equation}
where
\begin{equation*}
	\bm{a}=\begin{pmatrix}a\\a^\dagger\end{pmatrix}\quad,\quad \textbf{A}=\begin{pmatrix}0&\varepsilon\\ \varepsilon^*&0\end{pmatrix}\quad,\quad\bm{\Gamma}=\begin{pmatrix}\Gamma \\ \Gamma^\dagger\end{pmatrix}\;.
\end{equation*}
Eq. (\ref{eq:MLangevin}) is easiest solved in frequency domain, and by the Fourier transform,
\begin{equation*}\begin{aligned}
	a(t)&=\frac{1}{\sqrt{2\pi}}\int_{-\infty}^\infty\tilde{a}(\omega)e^{-i\omega t}d\omega\;,\\
	a^\dagger(t)&=\frac{1}{\sqrt{2\pi}}\int_{-\infty}^\infty\tilde{a}^\dagger(\omega)e^{i\omega t}d\omega=\frac{1}{\sqrt{2\pi}}\int_{-\infty}^\infty\tilde{a}^\dagger(-\omega)e^{-i\omega t}d\omega\;,
\end{aligned}\end{equation*}
eq. (\ref{eq:MLangevin}) becomes
\begin{equation}\label{eq:FMLangevin}
	-i\omega\tilde{\bm{a}}=\left(\textbf{A}-\gamma\bm{1}\right)\tilde{\bm{a}}+\tilde{\bm{\Gamma}}\;,
\end{equation}
\begin{equation*}
	\tilde{\bm{a}}=\begin{pmatrix} \tilde{a}(\omega) \\ \tilde a^\dagger(-\omega)\end{pmatrix} \quad,\quad \tilde{\bm{\Gamma}}=\begin{pmatrix} \tilde{\Gamma}(\omega) \\ \tilde \Gamma^\dagger(-\omega)\end{pmatrix}\;.
\end{equation*}
Solving for $\tilde{\bm{a}}$ in (\ref{eq:FMLangevin}),
\begin{equation*}
	\tilde{\bm{a}}=-\left(\textbf{A}+(i\omega-\gamma)\bm{1}\right)^{-1}\tilde{\bm{\Gamma}}\;,
\end{equation*}
leads to the solution
\begin{equation}\begin{aligned}\label{eq:SolLangevin}
	\tilde{a}(\omega)&=\frac{(i\omega-\gamma)\tilde{\Gamma}(\omega)-\varepsilon\tilde{\Gamma}^\dagger(-\omega)}{|\varepsilon|^2-(i\omega-\gamma)^2}\;,\\
	\tilde{a}^\dagger(-\omega)&=\frac{(i\omega-\gamma)\tilde{\Gamma}^\dagger(-\omega)-\varepsilon\tilde{\Gamma}(\omega)}{|\varepsilon|^2-(i\omega-\gamma)^2}\;,
\end{aligned}\end{equation}
of the annihilation and creation operator of the cavity frequency modes.

\begin{figure}
	\centering
	\includegraphics[width=5cm]{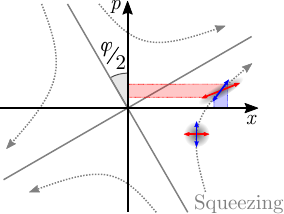}
	\caption{\label{fig:squeezing}Illustration of the phase space transformation under squeezing with arbitrary $\phi$ (grey). Red illustrates how noise initially in the $x$-quadrature is mixed into the $p$-quadrature when squeezing, and similar (by blue) how noise in the $p$-quadrature is mixed into the $x$-quadrature.}
\end{figure}

With the solution of the Langevin equations in (\ref{eq:SolLangevin}), and writing the effective pump intensity as $\varepsilon=|\varepsilon|e^{i\phi}$, the cavity mode quadrature in frequency domain, $\tilde{q}_a(\theta)$, becomes
\begin{equation*}\begin{aligned}
	\tilde{q}_a(\theta)&=\frac{1}{\sqrt{2}}\left(e^{-i\theta}\tilde{a}(\omega)+e^{i\theta}\tilde{a}^\dagger(-\omega)\right)\\
	&=\frac{1}{\sqrt{2}}\frac{
	e^{-i\theta}\left[(i\omega-\gamma)\tilde{\Gamma}(\omega)-|\varepsilon|e^{i\phi}\tilde{\Gamma}^\dagger(-\omega)\right]+e^{i\theta}\left[ (i\omega-\gamma)\tilde{\Gamma}^\dagger(-\omega)-|\varepsilon|e^{-i\phi}\tilde{\Gamma}(\omega)\right]
	}{|\varepsilon|^2-(i\omega-\gamma)^2}\\
	&=\frac{
	(i\omega-\gamma)\frac{1}{\sqrt{2}}\left[e^{-i\theta}\tilde{\Gamma}(\omega)+e^{i\theta}\tilde{\Gamma}(-\omega)\right]-|\varepsilon|\frac{1}{\sqrt{2}}\left[e^{-i(\phi-\theta)}\tilde{\Gamma}(\omega)+e^{i(\phi-\theta)}\tilde{\Gamma}^\dagger(-\omega)\right]
	}{|\varepsilon|^2-(i\omega-\gamma)^2}\\
	&=\frac{(i\omega-\gamma)\tilde{q}_\Gamma(\theta)-|\varepsilon|\tilde{q}_\Gamma(\phi-\theta)}{|\varepsilon|^2-(i\omega-\gamma)^2}\;.
\end{aligned}\end{equation*}
Here,
\begin{equation*}\begin{aligned}
	\tilde{q}_\Gamma(\theta)&=\frac{1}{\sqrt{2}}\left(e^{-i\theta}\tilde{\Gamma}(\omega)+e^{i\theta}\tilde{\Gamma}^\dagger(-\omega)\right)\\
	&=\frac{1}{\sqrt{2}}\left( e^{-i\theta}\left[\sqrt{2\gamma_c}\tilde{a}_\text{in}(\omega) + \sqrt{2\gamma_l}\tilde{a}_l(\omega) + \sqrt{2\gamma_s}\tilde{a}_s(\omega)\right] \right.\\
	&\hspace{3cm} \left.+ e^{i\theta}\left[\sqrt{2\gamma_c}\tilde{a}^\dagger_\text{in}(-\omega) + \sqrt{2\gamma_l}\tilde{a}^\dagger_l(-\omega) + \sqrt{2\gamma_s}\tilde{a}^\dagger_s(-\omega)\right]\right)\\
	&=\sqrt{2\gamma_c}\tilde{q}_\text{in}(\theta)+\sqrt{2\gamma_l}\tilde{q}_l(\theta)+\sqrt{2\gamma_s}\tilde{q}_s(\theta)
\end{aligned}\end{equation*}
is a weighted sum of quadratures in the modes of the noise operator $\Gamma$. Writing the noise operator quadrature in terms of amplitude and phase quadrature, $\tilde{q}_\Gamma(\theta)=\tilde{x}_\Gamma\cos\theta+\tilde{p}_\Gamma\sin\theta$, the amplitude and phase quadrature of the cavity mode becomes
\begin{equation}\label{eq:x}
	\tilde{x}_a=\tilde{q}_a(0)=\frac{(i\omega-\gamma)\tilde{q}_\Gamma(0)-|\varepsilon|\tilde{q}_\Gamma(\phi)}{|\varepsilon|^2-(i\omega-\gamma)^2}=\frac{(i\omega-\gamma-|\varepsilon|\cos\phi)\tilde{x}_\Gamma-|\varepsilon|\sin\phi\tilde{p}_\Gamma}{|\varepsilon|^2-(i\omega-\gamma)^2}\;
\end{equation}
\begin{equation}\label{eq:p}
	\tilde{p}_a=\tilde{q}_a(\pi/2)=\frac{(i\omega-\gamma)\tilde{q}_\Gamma(\pi/2)-|\varepsilon|\tilde{q}_\Gamma(\phi-\pi/2)}{|\varepsilon|^2-(i\omega-\gamma)^2}=\frac{(i\omega-\gamma+|\varepsilon|\cos\phi)\tilde{p}_\Gamma-|\varepsilon|\sin\phi\tilde{x}_\Gamma}{|\varepsilon|^2-(i\omega-\gamma)^2}\;,
\end{equation}
where it is used that $\tilde{q}_\Gamma(\phi-\pi/2)=\tilde{x}_\Gamma\sin\phi-\tilde{p}_\Gamma\cos\phi$. In conclusion, when $\phi=0$ or $\pi$ the cavity mode amplitude quadrature does not depend on the external noise phase quadrature, and similarly, the cavity phase quadrature does not depend on the external noise amplitude quadrature. This is expected, as for $\phi=0$ and $\pi$ we have squeezing exactly along the $p$- and $x$-quadrature respectively. However, when that is not the case ($\phi\neq0,\pi$), the quadratures mix as the phase space is stretched in some direction not along $x$ or $p$. This is illustrated in Figure \ref{fig:squeezing}.

In the experimental setup, we have squeezing in the $p$-quadrature, $\phi=\pi$, and (\ref{eq:x}) and (\ref{eq:p}) simplify to
\begin{equation*}
	\tilde{x}_a=\frac{i\omega-\gamma+|\varepsilon|}{|\varepsilon|^2-(i\omega-\gamma)^2}\tilde{x}_\Gamma=\frac{\tilde{x}_\Gamma}{\gamma+\varepsilon-i\omega}\;,
\end{equation*}
\begin{equation*}
	\tilde{p}_a=\frac{i\omega-\gamma-|\varepsilon|}{|\varepsilon|^2-(i\omega-\gamma)^2}\tilde{p}_\Gamma=\frac{\tilde{x}_\Gamma}{\gamma-\varepsilon-i\omega}\;.
\end{equation*}
The output field mode, $a_\text{out}$, is related to the cavity mode by $a_\text{out}=\sqrt{2\gamma_c}a-a_\text{in}$ \cite{collett84}, and the quadrature of the electric field coupled out of the cavity through the coupling mirror becomes
\begin{equation*}
	\tilde{x}_\text{out}=\sqrt{2\gamma_c}\frac{\tilde{x}_\Gamma}{\gamma+\varepsilon-i\omega}-\tilde{x}_\text{in}=\frac{\gamma_c-\gamma_l-\gamma_s-\varepsilon+i\omega}{\gamma+\varepsilon-i\omega}\tilde{x}_\text{in}+\frac{2\sqrt{\gamma_c\gamma_l}}{\gamma+\varepsilon-i\omega}\tilde{x}_l+\frac{2\sqrt{\gamma_c\gamma_s}}{\gamma+\varepsilon-i\omega}\tilde{x}_s\;,
\end{equation*}
\begin{equation*}
	\tilde{p}_\text{out}=\sqrt{2\gamma_c}\frac{\tilde{p}_\Gamma}{\gamma+\varepsilon-i\omega}-\tilde{p}_\text{in}=\frac{\gamma_c-\gamma_l-\gamma_s+\varepsilon+i\omega}{\gamma-\varepsilon-i\omega}\tilde{p}_\text{in}+\frac{2\sqrt{\gamma_c\gamma_l}}{\gamma-\varepsilon-i\omega}\tilde{p}_l+\frac{2\sqrt{\gamma_c\gamma_s}}{\gamma-\varepsilon-i\omega}\tilde{p}_s\;.
\end{equation*}
Finally, when the cavity output field is detected by homodyne detection, where $x_\text{out}$ and $p_\text{out}$ are measured in the rotating frame, it is the photo current we measure. Here, the photo current is a direct measure of $x_\text{out}$ and $p_\text{out}$, and the photo current power spectrum is proportional to the spectral density $S_q(\omega)$ of the quadrature $q(\theta)$ defined by $S_q(\omega)\delta(\omega+\omega')=\braket{\tilde{q}(\omega)\tilde{q}(\omega')}$ \cite{mandel95}. Using that $\braket{q_iq_j}=\braket{q_i}\braket{q_j}$ for independent modes $i$ and $j$, and that the only input mode to the cavity which is not simply vacuum is the seed, i.e. $\braket{q_\text{in}}=\braket{q_l}=0$ and $\braket{\Delta q_\text{in}^2}=\braket{\Delta q_l^2}=1/2$ for $q=x,p$, the spectral density of $x_\text{out}$ and $p_\text{out}$ becomes, after some cumbersome simplification,
\begin{equation}\label{eq:xout}
	S_x(\omega)=\braket{\tilde{x}_\text{out}(\omega)\tilde{x}_\text{out}(-\omega)}=\frac{1}{2}-\frac{2\gamma_c\varepsilon}{(\gamma+\varepsilon)^2+\omega^2}+\frac{4\gamma_c\gamma_s}{(\gamma+\varepsilon)^2+\omega^2}\left(\braket{\Delta x_s^2}-\frac{1}{2}\right)\;,
\end{equation}
\begin{equation}\label{eq:pout}
	S_p(\omega)=\braket{\tilde{p}_\text{out}(\omega)\tilde{p}_\text{out}(-\omega)}=\frac{1}{2}+\frac{2\gamma_c\varepsilon}{(\gamma-\varepsilon)^2+\omega^2}+\frac{4\gamma_c\gamma_s}{(\gamma-\varepsilon)^2+\omega^2}\left(\braket{\Delta p_s^2}-\frac{1}{2}\right)\;.
\end{equation}
In conclusion, for the seed beam simply being vacuum or a pure coherent state, i.e. $\braket{\Delta x_s^2}=\braket{\Delta p_s^2}=1/2$, (\ref{eq:xout}) and (\ref{eq:pout}) reduces to the well known squeezing spectra, where the first term is the vacuum noise, to which the second term add or subtract to form an anti-squeezed or squeezed spectrum respectively. However, for seed beam noise larger than the vacuum noise, the third term in (\ref{eq:xout}) and (\ref{eq:pout}) becomes positive and adds to the squeezing spectra. To encounter any loss in experimental setup, the decay rate $\gamma_c$ in the numerators of the two last terms in (\ref{eq:xout}) and (\ref{eq:pout}) can be replaced by $\eta\gamma$, where $\eta$ is the total efficiency of the experimental setup, ranging from the OPO escape efficiency, $\gamma_c/\gamma$, to the homodyne detection efficiency.

In the experimental setup, even with the mirror through which the seed beam is coupled into the cavity being a high reflector ($\gamma_s$ being small), the seed beam noise is large enough for the third term in (\ref{eq:xout}) and (\ref{eq:pout}) to be visible. To take this into account when predicting the squeezing spectra to the, we need to estimate $\gamma_s$ and $\braket{\Delta q_s^2}$ for $q=x,p$. To do this the spectral density of $x$ and $p$ are measured without pump power in the cavity ($\varepsilon=0$), $S_q^0(\omega)$, leading to 
\begin{equation*}
	S_q^0(\omega)=\frac{1}{2}+4\frac{4\eta\gamma\gamma_s}{\gamma^2+\omega^2}\left(\braket{\Delta q_s^2}-\frac{1}{2}\right)=\frac{1}{2}+\frac{K_q}{\gamma^2+\omega}\quad,\quad q=x,p
\end{equation*}
\begin{equation*}
	\Updownarrow
\end{equation*}
\begin{equation}\label{eq:Kq}
	K_q=4\eta\gamma\gamma_s\left(\braket{\Delta q_s^2}-\frac{1}{2}\right)=(\gamma^2+\omega^2)\left(S_q^0(\omega)-\frac{1}{2}\right)\;,
\end{equation}
where the total efficiency, $\eta$, of the setup is included.
Here, since $\gamma_s\ll\gamma_c,\gamma_l$,
\begin{equation*}
	\gamma=\gamma_c+\gamma_l+\gamma_s\approx\gamma_c+\gamma_l=\frac{1-\sqrt{1-T_c}}{\tau}+\frac{1-\sqrt{1-\mathcal{L}}}{\tau}=2\pi\SI{8.1}{MHz}
\end{equation*}
with $T_c=10\%$ coupling mirror transmission, $\mathcal{L}=0.55\%$ intracavity losses and $\tau=\SI{320}{mm}/(\SI{3E8}{m/s})$ OPO round trip time. Thus by measuring $S_q^0(\omega)$ and $\gamma$ we get $K_q$ from (\ref{eq:Kq}), we predict the squeezed and anti-squeezed spectra as
\begin{equation*}
	S_q(\omega)=\frac{1}{2}\mp\frac{2\varepsilon\gamma\eta}{(\gamma\pm\varepsilon)^2+\omega^2}+\frac{K_q}{(\gamma\pm\varepsilon)^2+\omega^2}\quad,\quad q=x,p\;.
\end{equation*}

\section{Bound on covariances}\label{sec:bound}
From measurements we got the covariance matrix for the operators $\bm{\xi} = (x_A,p_A,x_B,p_B)^T$ to be
\begin{equation}\label{eq:CM}
	\bm{\gamma}=V_0\begin{pmatrix}
		4.36 & a & -3.84 & 0.36 \\
		a & 4.43 & 0.45 & 3.92 \\
		-3.84 & 0.45 & 4.17 & b \\
		0.36 & 3.92 & b & 4.26
	\end{pmatrix}\;,
\end{equation}
where $V_0=1/2$ is the vacuum variance.

In the experiment, we aim at setting the phase angles such that $a=b=0$, but since $a$ and $b$ are not directly measured they could in principle deviate from zero. In the following we first estimate the allowed range of values for $a$ and $b$ using the uncertainty relation and given the actual measured entries of the covariance matrix. Second, we investigate how a phase off-set of the individually squeezed beams will lead to non-zero values of $a$ and $b$.  

To determine the bounds of $a$ and $b$ for $\bm{\gamma}$ to be physical, we use the covariance matrix uncertainty relation \cite{Simon2000}
\begin{equation}\label{eq:CMC}
	\bm{\gamma}+\frac{i}{2}\bm{\Omega}\geq 0\;,
\end{equation}
where the elements of $\bm{\Omega}$ are given by the commutator relations $[\xi_i, \xi_j] = i \Omega_{ij}$, hence
\begin{equation}
	\bm{\Omega}=
	\begin{pmatrix}
		0 & 1 & 0 & 0 \\
		-1 & 0 & 0 & 0 \\
		0 & 0 & 0 & 1 \\
		0 & 0 & -1 & 0
	\end{pmatrix} .
\end{equation}

\begin{figure}[h]
	\centering
	\includegraphics[width=0.9\textwidth]{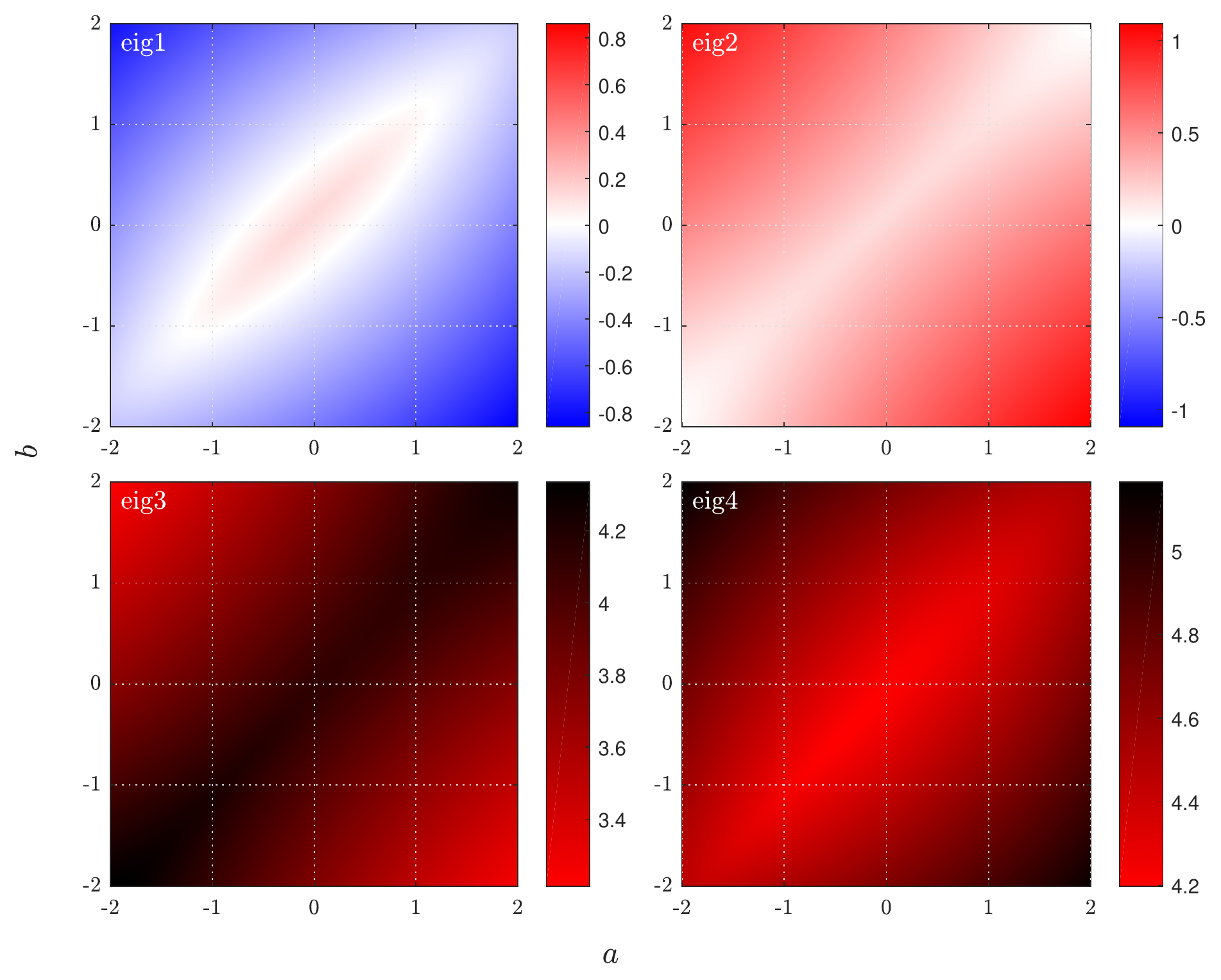}
	\caption{Numerical calculations of the eigenvalues of $\bm{\gamma}+\frac{i}{2}\bm{\Omega}$ for the covariance matrix shown in (\ref{eq:CM}).}
	\label{fig:eig}
\end{figure}

From \eqref{eq:CMC}, all eigenvalues of $\bm{\gamma}+\frac{i}{2}\bm{\Omega}$ must be non-negative.
In Figure \ref{fig:eig} the 4 eigenvalues of $\bm{\gamma}+\frac{i}{2}\bm{\Omega}$ for $a$ and $b$ in the range of $-2$ to $2$ is shown. In this interval only the first eigenvalue is seen to limit $a$ and $b$ in (\ref{eq:CMC}). The bounds of $a$ and $b$ are seen not to be independent, but they both share an absolute lower bound of zero (corresponding to the ideal case of $x$ and $p$ being completely uncorrelated). The lower and upper bound of $a$ (depending on $b$) and $b$ (depending on $a$) is
\begin{equation}
	a\in[-1.24,1.17]\;\;\;
\end{equation}
\begin{equation}
	b\in[-1.10,1.21]\;.
\end{equation}

\begin{figure}
    \centering
    \includegraphics[width=0.6\textwidth]{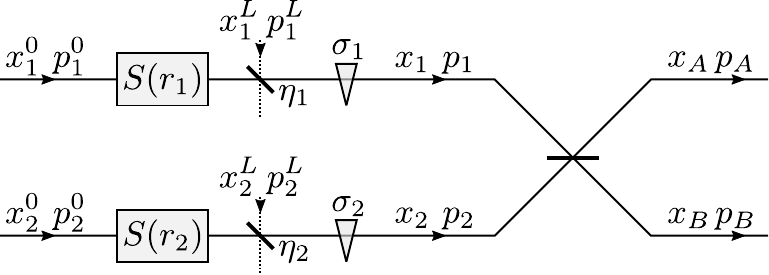}
    \caption{Two modes of vacuum, $x_1^0,p_1^0$ and $x_2^0,p_2^0$, are squeezed by $S(r_1)$ and $S(r_2)$ before they are subjected to loss ($1-\eta_1$ and $1-\eta_2$) and some phase change ($\sigma_1$ and $\sigma_2$). Finally they are interfered on a beam splitter to form two-mode squeezing. Here, mode 1 and 2 corresponds to the direct and delay path in the experimental setup. Notice, in the experimental setup the two modes 1 and 2 share the same temporal multiplexed squeezing source, such that the squeezing parameters $r_1=r_2$.}
    \label{fig:model}
\end{figure}

It is thus clear that by using a bona fide criterion for the covariance matrix, the range of possible xp-covarainces is quite large. In the following, we instead estimate the potential values for $a$ and $b$ by propagating a phase off-set of the input squeezed state through the system. We consider the setup in Figure \ref{fig:model}. Two squeezed states undergo a phase rotation, $\sigma$, and loss, $1-\eta$, and interfere on a balanced beam splitter 90$^\circ$ out of phase to produce a two-mode squeezed state. If the phase off-set is zero, the xp-covariances will be exactly zero. However, for a phase rotation, the quadratures are transformed as 
\begin{eqnarray*}
	x_1=\left(\sqrt{\eta_1} e^{-r_1}x_1^0-\sqrt{1-\eta_1}x_1^L\right)\cos{\sigma_1}-\left(\sqrt{\eta_1}e^{r_1}p_1^0+\sqrt{1-\eta_1}p_1^L\right)\sin{\sigma_1}\\
	p_1=\left(\sqrt{\eta_1} e^{-r_1}x_1^0-\sqrt{1-\eta_1}x_1^L\right)\sin{\sigma_1}+\left(\sqrt{\eta_1}e^{r_1}p_1^0+\sqrt{1-\eta_1}p_1^L\right)\cos{\sigma_1}\\
	x_2=\left(\sqrt{\eta_2} e^{r_2}x_2^0-\sqrt{1-\eta_2}x_2^L\right)\cos{\sigma_2}-\left(\sqrt{\eta_2}e^{-r_2}p_2^0+\sqrt{1-\eta_2}p_2^L\right)\sin{\sigma_2}\\
	p_2=\left(\sqrt{\eta_2} e^{r_2}x_2^0-\sqrt{1-\eta_2}x_2^L\right)\sin{\sigma_2}+\left(\sqrt{\eta_2}e^{-r_2}p_2^0+\sqrt{1-\eta_2}p_2^L\right)\cos{\sigma_2}
\end{eqnarray*}
and with $x_A=\frac{1}{\sqrt{2}}(x_1+x_2),\; x_B=\frac{1}{\sqrt{2}}(x_1-x_2),\; p_A=\frac{1}{\sqrt{2}}(p_1+p_2),\; p_B=\frac{1}{\sqrt{2}}(p_1-p_2)$, we find the correlation
\begin{equation*}
	C_{x_Ap_B}=\frac{1}{2}\left(\langle x_1p_1\rangle-\langle x_2p_2\rangle\right)=\frac{1}{2}\eta_1(e^{-2r_1}-e^{2r_1})V_0\cos{\sigma_1}\sin{\sigma_1}+\frac{1}{2}\eta_2(e^{-2r_2}-e^{2r_2})V_0\cos{\sigma_2}\sin{\sigma_2}
\end{equation*}
It is thus clear that a phase off-set of the individually squeezed modes results in non-zero xp-covariances. Interestingly, the covariances are identical to the intra-mode covariances:
\begin{equation*}
	C_{x_Ap_A}=C_{x_Bp_B}=C_{x_Ap_B}=C_{x_Bp_A}
\end{equation*}
which means that the expected off-set values of $a$ and $b$ should be similar to the measured values for $C_{x_Ap_A}$ and $C_{x_Bp_B}$ under the above mentioned assumptions. We have now seen how a phase off-set will give rise to non-zero covariances. It is however important to note that symmetric phase diffusion noise centered around the perfectly aligned phase will not produce non-zero values of the covariances.     

